\documentclass[amsfonts,amsmath,amssymb,aps,pra,showpacs,showkeys,reprint]{revtex4-1}

\usepackage{graphicx}
\usepackage[english]{babel}
\usepackage{bm}

\begin{document}
\title{Testing the neutrality of matter by acoustic means in a spherical resonator}

\author{G. Bressi}\thanks{deceased}
\affiliation{INFN, Sez.\ di Pavia, Via U.\ Bassi 6, I-27100 Pavia, Italy}
\author{G. Carugno}
\affiliation{INFN, Sez.\ di Padova, Via F.\ Marzolo 8, I-35131 Padova, Italy}
\author{F. \surname{Della Valle}}\email{federico.dellavalle@ts.infn.it}
\affiliation{Dip.\ di Fisica, Univ.\ di Trieste and INFN, Sez.\ di Trieste, Via A.\ Valerio 2, I-34127 Trieste, Italy}
\author{G. Galeazzi}
\author{G. Ruoso}
\affiliation{INFN, Laboratori Nazionali di Legnaro, Viale dell'Universit\`a 2, I-35020 Legnaro (PD), Italy}
\author{G. Sartori}
\affiliation{INFN, Sez.\ di Padova and Dip.\ di Fisica, Univ.\ di Padova, Via F.\ Marzolo 8, I-35131 Padova, Italy}


\begin{abstract}
New measurements to test the neutrality of matter by acoustic means are reported. The apparatus is based on a spherical capacitor filled with gaseous SF$_6$ excited by an oscillating electric field. The apparatus has been calibrated measuring the electric polarizability. Assuming charge conservation in the $\beta$ decay of the neutron, the experiment gives a limit of $\epsilon_\text{p-e}\lesssim1\times10^{-21}$ for the electron-proton charge difference, the same limit holding for the charge of the neutron. Previous measurements are critically reviewed and found incorrect: the present result is the best limit obtained with this technique.
\end{abstract}


\maketitle

\section{Introduction}
The most recent experimental value of the elementary charge $e$ is \cite{CODATA2006}
\[e=1.602\:176\:487(40)\times10^{-19}\text{~C}\]
where $0.000\,000\,040\times10^{-19}$~C is the standard uncertainty, equivalent to $2.5\times10^{-8}$ relative standard uncertainty. It is commonly accepted that the charges of the electron and of the proton have equal magnitude, and this fact is usually associated with the principles of charge conservation and of baryon and lepton conservation, as well as with the symmetry between particles and antiparticles. However, as the charge symmetry lacks a corresponding conservation law \cite{Feinberg1959}, there is no compelling requirement for its validity: it rests uniquely on experimental bases. Nevertheless, and quite obviously, the structure of the elementary charges is of paramount importance for theoretical physics since 1931, when Dirac pointed out that the existence of magnetic monopoles would imply charge quantization \cite{Dirac1931}. The uniqueness of the values of the elementary charges for both leptons and baryons implies a novel connection, beyond Standard Model, between the two families of particles, suggesting, for example, the existence of hypothetical lepto-quark particles which decay in both leptons and quarks \cite{Okun1984}.

In the last century, several measurements have been performed in order to test the symmetry of the elementary charges, using different laboratory approaches; to this purpose, most experiments probe the neutrality of common matter. The first laboratory method used a gas flowing out of a container whose electrostatic potential could be measured \cite{Piccard1925,Hillas1959,King1960}. These measurements were initially spurred by Einstein's suggestion that a small difference between electron and proton charges could account for the magnetic field of the Earth \cite{Piccard1925}; it was noted also that a small deviation from the neutrality of matter could explain the expansion of the universe \cite{Hughes1949,Lyttleton1959}; these phenomena could also be due to a nonzero neutron charge \cite{Hillas1959}. By now, both hypotheses have been ruled out by experiment. In the beam deflection method, the electric properties of a single species are tested: measurements are performed on molecules \cite{Hughes1949,Zorn1960,Zorn1963,Fraser1968,Hughes1988} and neutrons \cite{Shapiro1956,Shull1967,Gahler1982,Baumann1988}. A modern version of Millikan's experiment applies an electrostatic force to magnetically suspended small bodies \cite{Trischka1960}; this method was initially devised to detect the presence of free quarks in common matter, but it allows also to put limits on the neutrality of matter \cite{Stover1967,Gallinaro1977,Marinelli1984}. The latest laboratory method applies an alternating electric field to a gas contained inside an acoustic cavity; if the gas molecules carry an electric charge, a sound wave is generated. This method is an original idea by Dylla and King \cite{Dylla1973}, and is often cited as one of the best results on this topic \cite{PDG2010}. However, in the following we show that the results of the experiment were incorrect, as they were affected by a few errors. Among the planned experiments, we want to cite the proposed use of a torsion balance \cite{Unnikrishnan2004}, and of atom interferometry \cite{Arvanitaki2008}. Limits on charge asymmetry are cast also by model-dependent astrophysical methods \cite{Sengupta2000}. Moreover, one should mention a test on the charges of positrons and antiprotons, derived from measurements of their cyclotron resonance frequencies and from spectroscopic data \cite{Hughes1992}. Limits have been set also on the neutrality of the neutrino by astrophysical methods \cite{Bernstein1963}.

The results of all the experiments performed so far are consistent with the usual view that matter is neutral, and have pushed further and further the limits on charge asymmetry between proton and electron and on the charge of the neutron. The results are best described in terms of two parameters
\[\epsilon_\text{p-e}\equiv\frac{q_\text{p}+q_\text{e}}{e}\qquad\text{and}\qquad\epsilon_\text{n}\equiv\frac{q_\text{n}}{e},\]
where $q_\text{p}$, $q_\text{e}$ and $q_\text{n}$ are the electric charges of protons, electrons and neutrons, respectively. If a body containing $Z$ protons, $Z$ electrons and $N$ neutrons is measured to be neutral in an experiment capable of a sensitivity $\delta q$, one has
\[|Z\epsilon_\text{p-e}+N\epsilon_\text{n}|\,e\leq\delta q.\]
In order to disentangle the two contributions, one has to perform independent measurements on at least two systems with different $Z$ and/or $N$. A common approach assumes instead charge conservation in the $\beta$ decay of the neutron $n\rightarrow p+e^-+\overline{\nu}$ (and the neutrality of the antineutrino); with these hypotheses $\epsilon_\text{p-e}=\epsilon_\text{n}\equiv\epsilon_q$ and
\[|\epsilon_q|\leq\frac{\delta q}{(Z+N)e}\approx\frac{\delta q}{e}\,\frac{m_\text{p}}{m},\]
where $m_\text{p}$ and $m$ are respectively the masses of the proton and of the molecule or the sample.

\begingroup
\squeezetable
\begin{table}[htb]
\begin{center}
\begin{tabular}{|c|c|c|r|r|r|c|} \hline\hline
year & method & species & \multicolumn{1}{c|}{$Z$} & \multicolumn{1}{c|}{nucleons} & \multicolumn{1}{c|}{limit} & Ref.\\ \hline\hline
\multicolumn{7}{|c|}{$\epsilon_q$} \\ \hline
1925 & 1 & CO$_2$    &     22 &                     44 &   $5\times10^{-21}$ & \cite{Piccard1925}          \\
1959 & 1 & Ar        &     18 &                     40 &   $1\times10^{-21}$ & \cite{Hillas1959,Bondi1959} \\
1960 & 1 & H$_2$, He &   2, 2 &                   2, 4 &      $\sim10^{-20}$ & \cite{King1960}             \\
1973 & 1 & H$_2$     &      2 &                      2 & $3.6\times10^{-21}$ & \cite{King1973}             \\
1973 & 1 & He        &      2 &                      4 & $1.4\times10^{-21}$ & \cite{King1973}             \\
1973 & 1 & SF$_6$    &     70 &                    146 &   $3\times10^{-23}$ & \cite{King1973}             \\
1949 & 2 & CsI       &    108 &                    260 & $1.5\times10^{-15}$ & \cite{Hughes1949}           \\
1960 & 2 & CsF       &     64 &                    152 & $1.3\times10^{-16}$ & \cite{Zorn1960}             \\
1963 & 2 & Cs        &     55 &                    133 & $4.2\times10^{-19}$ & \cite{Zorn1963}             \\
1968 & 2 & Cs        &     55 &                    133 & $1.3\times10^{-20}$ & \cite{Fraser1968}           \\
1988 & 2 & Cs        &     55 &                    133 & $1.3\times10^{-20}$ & \cite{Hughes1988}           \\
1967 & 3 & Fe        &        & $\sim2.6\times10^{18}$ & $0.8\times10^{-19}$ & \cite{Stover1967}           \\
1977 & 3 & Fe        &        & $\sim1.2\times10^{20}$ &      $\sim10^{-20}$ & \cite{Gallinaro1977}        \\
1984 & 3 & steel     &        &   $\sim2\times10^{19}$ & $1.6\times10^{-21}$ & \cite{Marinelli1984}        \\
1973 & 4 & SF$_6$    &     70 &                    146 & $1.3\times10^{-21}$ & \cite{Dylla1973}            \\
\hline\hline
\multicolumn{7}{|c|}{$\epsilon_\text{p-e}$} \\ \hline
1959 & 1 & N$_2$, Ar & 14, 18 &                 28, 40 &   $4\times10^{-20}$ & \cite{Hillas1959,Bondi1959} \\
1963 & 2 & H$_2$     &      2 &                      2 &   $1\times10^{-15}$ & \cite{Zorn1963}             \\
1963 & 2 & K, Cs     & 19, 55 &                39, 133 & $2.7\times10^{-17}$ & \cite{Zorn1963}             \\
1968 & 2 & K, Cs     & 19, 55 &                39, 133 & $3.5\times10^{-19}$ & \cite{Fraser1968}           \\
1988 & 2 & K, Cs     & 19, 55 &                39, 133 &   $2\times10^{-19}$ & \cite{Hughes1988}           \\
\hline\hline
\multicolumn{7}{|c|}{$\epsilon_\text{n}$} \\ \hline
1959 & 1 & N$_2$, Ar & 14, 18 &                 28, 40 &   $4\times10^{-20}$ & \cite{Hillas1959,Bondi1959} \\
1956 & 2 & $n$       &      0 &                      1 & $0.6\times10^{-11}$ & \cite{Shapiro1956}          \\
1963 & 2 & K, Cs     & 19, 55 &                39, 133 &   $2\times10^{-17}$ & \cite{Zorn1963}             \\
1967 & 2 & $n$       &      0 &                      1 & $3.7\times10^{-18}$ & \cite{Shull1967}            \\
1968 & 2 & K, Cs     & 19, 55 &                39, 133 & $2.7\times10^{-19}$ & \cite{Fraser1968}           \\
1982 & 2 & $n$       &      0 &                      1 & $1.4\times10^{-20}$ & \cite{Gahler1982}           \\
1988 & 2 & $n$       &      0 &                      1 & $1.1\times10^{-21}$ & \cite{Baumann1988}          \\
1988 & 2 & K, Cs     & 19, 55 &                39, 133 & $1.5\times10^{-19}$ & \cite{Hughes1988}           \\
\hline\hline\end{tabular}
\caption{\label{NeuTab}One sigma upper limits on the neutrality of matter from direct laboratory measurements. Methods: 1, gas efflux; 2, beam deflection; 3, suspended body; 4, acoustic. All the experiments are consistent with the assumption that matter is neutral. The present work questions the quoted limit obtained by method 4.}
\end{center}
\end{table}
\endgroup

The actual experimental situation is summarized in Tab.~\ref{NeuTab}. The measurements performed with the gas efflux method, leaving aside Ref. \cite{King1973}, give a limit on $\epsilon_q$ of the order of $10^{-21}$, a value obtained from careful subtraction of spurious effects of uncertain origin. The beam deflection method has, for molecules, a sensitivity one order of magnitude worst than the previous one; it heavily relies on a modelization of the beam profile and energetics and suffers from electric field spatial inhomogeneities, which exert forces on neutral molecules; it is difficult to think to improve this experimental set-up. Also the suspended body method has reached a sensitivity of the order of $10^{-21}$, which might be improved by a factor 10 by increasing the statistics \cite{Marinelli1984}; gaining a larger factor relies on the possibility of using bigger bodies, which is not trivial \cite{Marinelli1984}. In order to try to improve the limits on charge asymmetry, we chose the experimental set-up of a gas filled electric capacitor used as an acoustic resonant cavity \cite{Dylla1973}, which seemed less sensitive to systematic artefacts and at the same time most passible of upgrades. We considered, in particular, that a better sensitivity could be reached {\em i}) by increasing the gas pressure and hence the quality factor $Q$ of the cavity, and {\em ii}) by improving the detection method. In fact, in the original experiment the pressure in the cavity was limited to 1~atm; moreover we judged that we would have benefitted by the use of more modern electronics. But there were also other reasons for repeating the 1973 experiment by Dylla and King: in the first place the authors did not realize that their resonant mode was almost frequency degenerate with another one, a situation that calls for a complicated analysis. Moreover, we show that their calculation of the neutrality signal was wrong by a factor about sixty, while the value of the calibration signal they quote was wrong by a factor about three. As a consequence, their limit on electron-proton charge difference is, in reality, not better than about $10^{-19}$.

In Sect.~II we develop a simple mathematical description of the electrically excited acoustic effects in a spherical resonator; Sect.~III is a detailed account of the experimental set-up and of the method; Sect.~IV-A describes the results of the polarization measurements used to calibrate the apparatus; neutrality results are discussed in Sect.~IV-B; conclusions follow.

\section{Principle of measurement}

\begin{figure}[htb]
\begin{center}
\includegraphics[width=4cm]{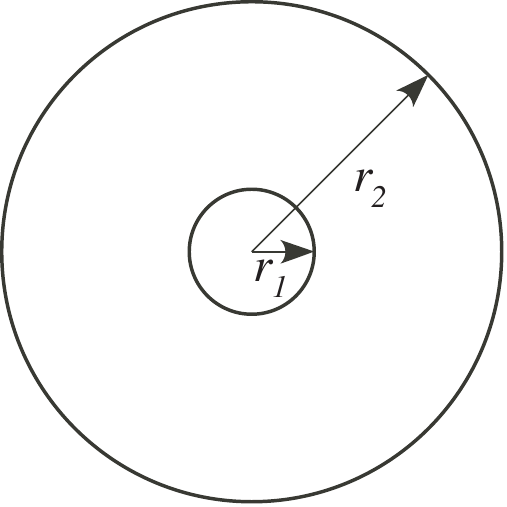}
\caption{\label{cavity}Scheme of a spherical acoustic resonator. Rigid walls at radiuses $r_1$ and $r_2$ enclose an oscillating fluid.}
\end{center}
\end{figure}

To measure the neutrality of matter with the acoustic method \cite{Dylla1973}, an oscillating voltage is applied to a gas filled spherical capacitor with internal and external radiuses $r_1$ and $r_2$, respectively -- see Fig.~\ref{cavity}. Forces are exerted on the molecules in the first instance through electric polarization, but also if matter is not neutral. In both cases, acoustic waves are generated, whose amplitude is measured by means of a sensor placed in gas at the external surface. In the following we show that the two mechanisms can be discriminated; hence the method is able to cast limits on the non neutrality of matter.

A detailed analysis of the paper by Dylla and King \cite{Dylla1973} revealed inconsistencies in their Eqs.~(2) and (3). We have independently solved the differential equation for the acoustic wave both analytically \cite{Cazzola} and numerically. The most general analytical solution which takes into full account the dissipative forces is too complicated to be included in the present work. Only a simplified version is presented below, and only for the neutrality problem. Anyway, all the results obtained from the complete, the simplified and the numerical solutions are found to fully coincide.

\subsection{Electric forces in gases}

The force per unit volume on an homogeneous dielectric with electric permittivity $\epsilon=\epsilon_0\epsilon_r$ and a charge density $\rho$ subject to an electric field $\bm{E}$ is given by \cite{Stratton1941}
\[\bm{f}=\rho\bm{E}+\frac{1}{2}\bm{\nabla}\left(E^2\rho_m\frac{\partial\epsilon}{\partial\rho_m}\right)\]
with $\rho_m$ the mass density. The second term is the force associated with electric polarization. The dielectric constant relates to mass density through the Clausius-Mossotti law
\[\frac{\epsilon_r-1}{\epsilon_r+2}=A\qquad\text{or}\qquad\epsilon_r-1=\frac{3A}{1-A}\]
with
\[A=\frac{n\alpha}{3\epsilon_0},\]
where $\alpha$ is the electric polarizability and $n=N_A\rho_m/M$ is the number of molecules per unit volume, with $M$ the molecular weight. This leads to
\[\rho_m\frac{\partial\epsilon}{\partial\rho_m}=\frac{\epsilon_0}{3}\,(\epsilon_r-1)(\epsilon_r+2).\]
If the gas has low density and bears a charge asymmetry, then
\begin{eqnarray}
\nonumber\bm{f}&=&(Z\epsilon_\text{p-e}+N\epsilon_\text{n})\,\frac{\rho_m}{Zm_\text{p}+Nm_\text{n}}\,e\bm{E}+\\*
&&+\frac{N_A}{2}\,\frac{\rho_m}{M}\,\alpha\,\bm{\nabla}E^2.
\label{forcedensity}
\end{eqnarray}

The two terms can be distinguished from their dependence on the electric field: for example, if the electric field is modulated at a frequency $\nu$, the first term is associated with the response of the gas at frequency $\nu$ and allows the study of the charge asymmetry, while the second term generates a signal at frequency $2\nu$ which can be used for calibration of the apparatus.

\subsection{Sound wave equation}

To first order in the vibration amplitude, and neglecting viscosity, the equation of motion of a small portion of fluid is
\begin{equation}
-\bm{\nabla}P+\bm{f}=\rho_{m0}\,\frac{\partial\bm{v}}{\partial t},
\label{NewtEq}
\end{equation}
where $P$ is the pressure, $\rho_{m0}$ the unperturbed mass density and $\bm{v}$ the fluid velocity. In the same approximation, conservation of mass is written as
\begin{equation}
\frac{\partial\rho_m}{\partial t}=-\rho_{m0}\bm{\nabla}\cdot\bm{v}.
\label{masscont}
\end{equation}
Since acoustic waves propagate with little or no heat exchange between adjacent regions of the fluid, the left-hand side of this equation can be transformed to
\[\frac{\partial\rho_m}{\partial t}=\left(\frac{\partial\rho_m}{\partial P}\right)_S\frac{\partial P}{\partial t}=k_S\,\rho_{m0}\,\frac{\partial P}{\partial t},\]
where the subscript $S$ indicates an isoentropic process and $k_S$ is the adiabatic compressibility modulus of the fluid. By taking the divergence of Eq.~(\ref{NewtEq}) and the time derivative of Eq.~(\ref{masscont}) and summing, we obtain an inhomogeneous scalar wave equation for the pressure $P$:
\begin{equation}
\nabla^2P-\frac{1}{c^2}\frac{\partial^2P}{\partial t^2}=\bm{\nabla}\cdot\bm{f}
\label{waveq}
\end{equation}
with wave speed $c$ given by
\[c^2=\left(\frac{\partial P}{\partial\rho_m}\right)_S=\frac{1}{k_S\,\rho_{m0}}.\]
If the fluid is an ideal gas, then the Laplace's equation holds
\begin{equation}
c^2=\frac{\gamma P_0}{\rho_{m0}}=\frac{\gamma RT}{M},
\label{wavespeed}
\end{equation}
where $\gamma$ is the ratio of the constant pressure to the constant volume specific heats of the gas. In this case wave speed scales as the square root of the absolute temperature $T$ of the gas and is independent of pressure. For a real gas, this is true only in the limit $P\rightarrow0$. At non-zero pressures, a more general theory predicts a dependence of $c$ on pressure \cite{Trusler1991}. Anyway, in the context of the present work, no deeper investigation on this subject is needed: changes of the value of $c$ will show as a rescaling of the frequency spectrum.

\subsection{Frequency spectrum of a spherical resonator}

In a bounded volume, the homogeneous Helmholtz equation associated with Eq.~(\ref{waveq}) determines a discrete frequency spectrum. In a spherical geometry the equation can be solved by variable separation. The angular part of the wave function is given by the spherical harmonics $Y_{lm}(\theta,\phi)$ \cite{Morse1953}, while the radial function $R(r)$ is a solution of
\begin{equation}
\displaystyle\frac{d}{d r}\left(r^{2}\frac{dR}{dr}\right)=[l(l+1)-k^2r^2]\,R,
\label{HelmEq}
\end{equation}
where $k=2\pi\nu/c=\omega/c$ is the wave vector. Solutions of this equation are the spherical Bessel functions of first and of second kind \cite{Morse1953}:
\begin{equation}
R_l(r)=A_l\,j_l(kr)+B_l\,y_l(kr)
\label{radialsol}
\end{equation}
with $A_l$ and $B_l$ arbitrary coefficients.

Let us suppose that the walls of the cavity at radiuses $r_1$ and $r_2$ are perfectly rigid. According to Eq.~(\ref{NewtEq}), on the cavity walls, where the radial component of the velocity is zero, the acoustic wave satisfies Neumann boundary conditions $R'_l(r_1)=R'_l(r_2)=0$, or
\begin{eqnarray*}
&A_l\,j'_l(kr_1)+B_l\,y'_l(kr_1)=0\\*
&A_l\,j'_l(kr_2)+B_l\,y'_l(kr_2)=0.
\end{eqnarray*}
If both equations must hold, then $j'_l(kr_1)\,y'_l(kr_2)-y'_l(kr_1)\,j'_l(kr_2)=0$. This last equation generates a discrete spectrum of allowed wave vectors $k_{nl}$, with index $n$ numbering the roots for a given $l$. The solution $k=0$ for $l=0$ is left aside for the purpose of the present work. The radial parts of the normal modes of a spherical acoustic cavity are then of the form
\[R_{nl}(r)=A'_{nl}\,\left[j_l(k_{nl}r)-\frac{j'_l(k_{nl}r_1)}{y'_l(k_{nl}r_1)}y_l(k_{nl}r)\right].\]
We call radial modes the spherically symmetrical ones with $l=0$, and angular modes the others. 

If $r_1=0$, namely the cavity is an empty sphere, the coefficient $B_l$ in Eq.~(\ref{radialsol}) is zero; Neumann boundary conditions can still be imposed and the set of allowed wave vectors changes accordingly \cite{Ferris1952}. In any case, at a single frequency $\omega_{nl}$ we have
\[P(\mathbf{r},\omega_{nl})=P_0+R_{nl}(r)\,e^{-i(\omega_{nl}t+\phi_{nl})}\,\sum_mA''_{nlm}Y_{lm}(\theta,\phi).\]

\begin{table}[htb]
\begin{center}
\begin{tabular}{||r||r|r|r||r|r|r||} \hline\hline
                 &  \multicolumn{3}{|c||}{$r_1=4$~cm} & \multicolumn{3}{|c||}{$r_1=6.35$~cm} \\ \hline
$l~\backslash~n$ &        1 &        2 &        3 &        1 &        2 &        3    \\ \hline
               0 & 23.4320 & 41.8903 & 60.8294 & 25.6396 & 47.6045 & 70.1426  \\ \hline
               1 & 10.2824 & 28.9789 & 45.5350 &  9.9725 & 29.2462 & 49.4758  \\ \hline
               2 & 16.6919 & 36.1174 & 52.0398 & 16.5510 & 35.2314 & 53.1542  \\ \hline
               3 & 22.5684 & 42.8426 & 59.3966 & 22.5258 & 42.0626 & 58.4603  \\ \hline
               4 & 28.2333 & 49.1890 &         & 28.2224 & 48.7960 & 64.9461  \\ \hline
               5 & 33.7823 & 55.3491 &         & 33.7797 & 55.1959 &          \\ \hline
               6 & 39.2554 & 61.3964 &         & 39.2548 & 61.3445 &          \\ \hline
               7 & 44.6742 &         &         & 44.6741 & 67.3440 &          \\ \hline
               8 & 50.0519 &         &         & 50.0518 &         &          \\ \hline
               9 & 55.3971 &         &         & 55.3971 &         &          \\ \hline
              10 & 60.7160 &         &         & 60.7160 &         &          \\ \hline
              11 &         &         &         & 66.0131 &         &          \\ \hline
              12 &         &         &         & 71.2917 &         &          \\ \hline
\hline\end{tabular}
\caption{\label{spher wavevec}Allowed wave vector values (in inverse meters) of acoustic waves inside a gas filled spherical resonator with $r_2=20$~cm, for $r_1=4$~cm and $r_1=6.35$~cm.}
\end{center}
\end{table}
\begin{table}[htb]
\begin{center}
\begin{tabular}{||r||r|r|r||} \hline\hline
$l~\backslash~n$ &        1 &        2 &        3 \\ \hline
               0 & 22.8505 & 39.4025 & 55.8203 \\ \hline
               1 & 10.5618 & 30.1062 & 46.5716 \\ \hline
               2 & 16.9649 & 37.0029 & 53.8668 \\ \hline
               3 & 22.9142 & 43.5723 & 60.7749 \\ \hline
               4 & 28.6635 & 49.9515 &          \\ \hline
               5 & 34.2967 & 56.1939 &          \\ \hline
               6 & 39.8532 & 62.3316 &          \\ \hline
               7 & 45.3545 &          &          \\ \hline
               8 & 50.8141 &          &          \\ \hline
               9 & 56.2407 &          &          \\ \hline
              10 & 61.6406 &          &          \\ \hline
\hline\end{tabular}
\caption{\label{Dylla wavevec}Allowed wave vector values (in inverse meters) of acoustic waves inside the gas filled spherical resonator used by Dylla and King \cite{Dylla1973}, having $r_1=1.27$~cm and $r_2=19.7$~cm.}
\end{center}
\end{table}

Tab.~\ref{spher wavevec} lists the first allowed wave vectors for the two cavities used in the present work, both with external radius $r_2=20$~cm, and with the internal wall at $r_1=4$~cm and $r_1=6.35$~cm, respectively. In Tab.~\ref{Dylla wavevec} the allowed wave vectors of the apparatus used by Dylla and King \cite{Dylla1973} are shown. Note that for this last configuration the radial modes are almost degenerate with angular modes, what makes that cavity not quite suitable for the experiment.  In fact, the authors meant to tune the $(1,0)$ mode, which is overlapping the $(1,3)$ for a $Q$ factor $\approx 1000$.

\subsection{Acoustic effect of electric polarization}

\begin{table}[htb]
\begin{center}
\begin{tabular}{|r|r|r|r@{.}l|cr@{}l|r|} \hline\hline
species & \multicolumn{1}{|c|}{ $M$ } & \multicolumn{1}{|c|}{ $c$ } & \multicolumn{2}{|c|}{ $\alpha$ } \\
 & (g/mol) & (m/s) & \multicolumn{1}{r}{($10^{-40}$} & \multicolumn{1}{l|}{C$\cdot$m$^2$/V)} \\ \hline
SF$_6$ & 146 & 135 & 7 & 277 \\ \hline
Xe     & 131 & 170 & 4 & 499 \\ \hline
Kr     &  84 & 220 & 2 & 764 \\ \hline
Ar     &  40 & 320 & 1 & 826 \\ \hline
N$_2$  &  28 & 350 & 1 & 936 \\ \hline
\hline\end{tabular}
\caption{\label{physparam}Physical parameters used in the calculations. Sound speed refers to $T=298$~K and $P_0=1$~atm. Static polarizability values are taken from Ref.~\cite{CRC1995}.}
\end{center}
\end{table}
\begin{table}[htb]
\begin{center}
\begin{tabular}{|c|c|r|r|r|} \hline\hline
$r_1$ & species & \multicolumn{3}{|c|}{mode $(n,0)$} \\
$(\text{cm})$ && \multicolumn{1}{c}{$(1,0)$} & \multicolumn{1}{c}{$(2,0)$} & \multicolumn{1}{c|}{$(3,0)$} \\ \hline\hline
&  SF$_6$ &  503 &  900 & 1307 \\
&      Xe &  634 & 1133 & 1646 \\
4.00 & Kr &  820 & 1467 & 2130 \\
&      Ar & 1193 & 2133 & 3098 \\
&   N$_2$ & 1305 & 2333 & 3388 \\
\hline\hline
&  SF$_6$ &  551 & 1023 & 1507 \\
&      Xe &  694 & 1288 & 1898 \\
6.35 & Kr &  898 & 1667 & 2456 \\
&      Ar & 1306 & 2424 & 3572 \\
&   N$_2$ & 1428 & 2652 & 3907 \\
\hline\hline
\end{tabular}
\caption{\label{resfreq}Resonance frequencies (in hertz) of the first radial modes of two cavities with $r_2=20$~cm and $r_1=4$~cm and $r_1=6.35$~cm, filled with different gases. All the values refer to $P_0=1$~atm and $T=298$~K.}
\end{center}
\end{table}

We want now to determine the acoustic effect of electric polarization on a gas contained inside a spherical capacitor with perfectly rigid walls. In doing this we will consider only acoustic modes having spherical symmetry, namely radial modes with $l=0$. The parameters used in the calculations are summarized in Tab.~\ref{physparam}; Tab.~\ref{resfreq} lists the resonance frequencies of the first radial modes of the two cavities used in the present work filled with different gases.

For ideal gases, the divergence of the second term of the force density of Eq.~(\ref{forcedensity}) is
\begin{equation}
\bm{\nabla}\cdot\bm{f}=6\,\frac{P_0}{k_\text{B}T}\,\alpha\left(\frac{1}{4\pi\epsilon}\right)^2\frac{C^2V^2}{r^6}\equiv\,\alpha\,\beta\,\frac{V^2}{r^6},
\label{divforce}
\end{equation}
where
\[C=4\pi\epsilon\frac{r_1r_2}{r_2-r_1}\]
is the capacitance of the spherical capacitor, and $V$ is the applied voltage difference. The parameter $\beta$ depends on cavity geometry. For cavities with $r_2=20$~cm and $r_1=4$~cm or $r_1=6.35$~cm it has values:
\begin{eqnarray}
\nonumber&\beta_4=\displaystyle\frac{6P_0}{k_BT}\left[\frac{r_1r_2}{r_2-r_1}\right]^2=3.69\;10^{23}\text{~m}^{-1}\,\frac{P_0}{\text{1~atm}}\,\frac{298\text{~K}}{T}\\*
\label{beta}\\*
\nonumber&\beta_{6.35}=1.28\;10^{24}\text{~m}^{-1}\,\displaystyle\frac{P_0}{\text{1~atm}}\,\frac{298\text{~K}}{T}.
\end{eqnarray}

For harmonic excitations, the wave equation (\ref{waveq}) reduces to an inhomogeneous scalar Helmholtz equation whose solution is obtained by numerical integration with boundary condition $dP/dr=f_r$ both at $r=r_1$ and at $r=r_2$. In the forcing term given by Eq.~(\ref{divforce}) the relevant quantity is the Fourier component of $V^2$ at a resonance frequency $\omega_{n0}$, which we call $V^2(\omega_{n0})$. In order to take into account the viscosity, we assume a complex wave vector:
\begin{equation}
k^2=\frac{\omega^2}{c^2}\left(1+\frac{i}{\pi\nu\tau}\right).
\label{complex_k}
\end{equation}
The relaxation time $\tau$ depends on base pressure, resonance mode and cavity; it is related to the quality factor through $Q=\pi\nu_{n0}\tau$. No attempt has been made to deduce $\tau$ from known values of viscosity, it is instead measured independently.

\begin{table}[htb]
\begin{center}
\begin{tabular}{|c|c|c|r|r|r|} \hline\hline
$r_1$ & value & species & \multicolumn{3}{|c|}{mode $(n,0)$} \\
(cm) &&& \multicolumn{1}{c}{$(1,0)$} & \multicolumn{1}{c}{$(2,0)$} & \multicolumn{1}{c|}{$(3,0)$} \\ \hline\hline
& $B_{n0}$~(m$^{-4}$) && 1100 & 956 & 756 \\
\hline
&& SF$_6$ & 295 & 257 & 203 \\
&&     Xe & 183 & 159 & 126 \\
4.00 & $\displaystyle\frac{P(r_2)}{Q\,V^2(\omega_{n0})}$ & Kr & 112 & 98 & 77 \\
& (fPa/V$^2$) & Ar & 74 & 64 & 51 \\
&& N$_2$ & 79 & 68 & 54 \\
\hline\hline
& $B_{n0}$~(m$^{-4}$) && 452 & 277 & 184 \\
\hline
&& SF$_6$ & 421 & 258 & 171 \\
&&     Xe & 260 & 160 & 106 \\
6.35 & $\displaystyle\frac{P(r_2)}{Q\,V^2(\omega_{n0})}$ & Kr & 160 & 98 & 65 \\
& (fPa/V$^2$) & Ar & 106 & 65 & 43 \\
&& N$_2$ & 112 & 69 & 46 \\
\hline\hline\end{tabular}
\caption{\label{polres}Numerically calculated results of acoustic effects due to electric polarization in spherical cavities having $r_2=20$~cm, for $r_1=4$~cm and $r_1=6.35$~cm. Dynamic pressure values are calculated through Eq.~(\ref{polpress}). All the numbers refer to $P_0=1$~atm and $T=298$~K -- see Eq.~(\ref{beta}).}
\end{center}
\end{table}

The output of the calculation is the amplitude of the pressure wave $P(r,\nu)$ at given $r$ and $\nu$. The calculations are performed for different values of the frequency around the resonance frequencies $\nu_{n0}$ in order to draw the amplitude resonance curve. The results are studied as a function of the values of $\tau$ and of $V^2(\omega_{n0})$. From the analysis of the results it turns out that the dynamic pressure peak values at $r_2$ due to electric polarization can be written as
\begin{eqnarray}
\nonumber P(r_2,\nu_{n0})&=&B_{n0}\,\pi\nu_{n0}\,\tau\,\alpha\,\beta\,V^2(\omega_{n0})=\\*
&=&B_{n0}\,Q_{n0}\,\alpha\,\beta\,V^2(\omega_{n0}).
\label{polpress}
\end{eqnarray}
The results of the calculations for the cavities employed in the present work are summarized in Tab.~\ref{polres}. For $V\sim1$~kV and $Q\sim10^3$, expected values of the dynamic pressure are $P\sim1$~mPa.

A similar calculation performed for the apparatus of Ref.~\cite{Dylla1973} ($r_1=1.27$~cm, $r_2=19.7$~cm, SF$_6$ at $P_0\approx0.8$~atm, $V=3500$~V rms, $Q=1100$) gives $P(r_2,\nu_{10})=1.42$~mPa, a value which is about three times larger than the 4.68~mdyn/cm$^2$ that can be obtained from Eq.~(3) of the same reference. As a consequence, since they use the polarization measurements to perform an absolute calibration of their instrument, the ratio of the voltage output of the microphone to the expected pressure signal, $V_s/P_D$, given in Eq.~(4) of Ref.~\cite{Dylla1973}, is too large of the same factor three.

\subsection{Acoustic effect of non-neutrality of matter}

The acoustic effect of an hypothetical charge asymmetry of proton and electron can be tackled in a similar way as the polarization effect. Unlike that case, however, for the non neutrality of matter the solution of the wave equation (\ref{radialHelm}) can be obtained in a closed form. In the following we show a simplified version of the analytical solution.

The divergence of the first term of Eq.~(\ref{forcedensity}) is zero everywhere inside the capacitor. Hence the wave function satisfies the homogeneous wave equation (\ref{HelmEq}) which, for radial modes, is
\begin{equation}
\frac{1}{r^2}\frac{d}{d r}\left(r^{2}\frac{dP}{dr}\right)+k^2\,P=0
\label{radialHelm}
\end{equation}
with inhomogeneous boundary conditions:
\[\frac{dP}{dr_{1,2}}=\epsilon_{q}e\,\frac{\rho_{m0}}{m_\text{p}}\,\frac{1}{4\pi\epsilon}\frac{CV}{r_{1,2}^2}\equiv\epsilon_q\,M\,\delta\,\frac{V}{r_{1,2}^2},\]
where $M=\rho_{m0}RT/P_0$ is the molecular mass and we have assumed $\epsilon_\text{p-e}=\epsilon_\text{n}=\epsilon_q$. Here, again, the coefficient $\delta$ does not depend on the gas species, but on the geometry of the cavity. Its values for cavities with $r_2=20$~cm and $r_1=4$~cm or $r_1=6.35$~cm are:
\begin{eqnarray}
\nonumber&\delta_4=\displaystyle\frac{e}{m_\text{p}}\frac{P_0}{RT}\frac{r_1r_2}{r_2-r_1}=1.96\;10^8\;\frac{\text{C$\cdot$mol}}{\text{kg$\cdot$m}^2}\,\frac{P_0}{\text{1~atm}}\,\frac{298\text{~K}}{T}\\*
\label{delta}\\*
\nonumber&\delta_{6.35}=3.64\;10^8\;\displaystyle\frac{\text{C$\cdot$mol}}{\text{kg$\cdot$m}^2}\,\frac{P_0}{\text{1~atm}}\,\frac{298\text{~K}}{T}.
\end{eqnarray}
From Eq.~(\ref{radialsol})
\[P(r)=A\,j_0(kr)+B\,y_0(kr)=\frac{A\,\sin(kr)}{kr}+\frac{B\,\cos(kr)}{kr}.\]
The coefficients $A$ and $B$ are determined by the boundary conditions
\begin{eqnarray*}
A\,j_0'(kr_1)+B\,y_0'(kr_1)=\epsilon_q\,M\,\delta\,\frac{V}{r_1^2}\\*
A\,j_0'(kr_2)+B\,y_0'(kr_2)=\epsilon_q\,M\,\delta\,\frac{V}{r_2^2},
\end{eqnarray*}
which give
\begin{widetext}
\begin{eqnarray*}
&A=\displaystyle\epsilon_q\,M\,\delta\,V\,k\,\frac{\cos(kr_1)-\cos(kr_2)+kr_1\,\sin(kr_1)-kr_2\,\sin(kr_2)}{k\Delta r\,\cos(k\Delta r)-(1+k^2r_1r_2)\,\sin(k\Delta r)}\\*
&B=\displaystyle\epsilon_q\,M\,\delta\,V\,k\,\frac{kr_1\,\cos(kr_1)-kr_2\,\cos(kr_2)-\sin(kr_1)+\sin(kr_2)}{-k\Delta r\,\cos(k\Delta r)+(1+k^2r_1r_2)\,\sin(k\Delta r)}
\end{eqnarray*}
where $\Delta r=r_2-r_1$. In turn one obtains
\[P(r)=\epsilon_q\,M\,\delta\,V\,\frac{kr_1\,\cos[k(r-r_1)]-kr_2\,\cos[k(r_2-r)]+\sin[k(r-r_1)]+\sin[k(r_2-r)]}{r\left[k\Delta r\,\cos(k\Delta r)-(1+k^2r_1r_2)\,\sin(k\Delta r)\right]}.\]
\end{widetext}
When the wavevector approaches a resonance value $k_{n0}$, this expression diverges, as the denominator goes to zero. To take into account dissipation, we use the complex definition of $k$ derived from Eq.~(\ref{complex_k}) in the case of weak dissipation:
\[k\approx\frac{\omega}{c}\left(1+\frac{i}{2Q}\right).\]
We note that for the real and imaginary parts of the wavevector it holds $k_I\ll k_R$. The complex value of $k$ can be substituted into the expression of $P(r)$, whose real and imaginary parts are in phase and out of phase with respect to the forcing field. To obtain an estimate of the amplitude of $P(r)$ at resonance, we consider separately the numerator and the denominator of the expression of $P(r)$ as a function of $k$. Let's call them $N(k)$ and $D(k)$, respectively. At resonance we have
\[\frac{N(k_{n0}+ik_I)}{D(k_{n0}+ik_I)}\approx\frac{N(k_{n0})+ik_I\dot N(k_{n0})}{D(k_{n0})+ik_I\dot D(k_{n0})}\approx\frac{2Q\,N(k_{n0})}{ik_{n0}\,\dot D(k_{n0})}\]
hence
\begin{widetext}
\begin{equation}\frac{P(r_2,\nu_{n0})}{\epsilon_q\,Q\,V(\omega_{n0})}=2M\,\delta\,\frac{k_{n0}r_1\,\cos(k_{n0}\Delta r)-k_{n0}r_2+\sin(k_{n0}\Delta r)}{k_{n0}^2r_2[-k_{n0}r_1r_2\Delta r\,\cos(k_{n0}\Delta r)-(r_1^2+r_2^2)\,\sin(k_{n0}\Delta r)]}.\end{equation}
\end{widetext}
This result is incompatible (analitically and numerically) with Eq.~(2) of Ref.~\cite{Dylla1973}.

\begin{table}[htb]
\begin{center}
\begin{tabular}{|c|c|c|r|r|r|} \hline\hline
$r_1$ & value & species & \multicolumn{3}{|c|}{mode $(n,0)$} \\
$(\text{cm})$ &&& \multicolumn{1}{|c}{$(1,0)$} & \multicolumn{1}{c}{$(2,0)$} & \multicolumn{1}{c|}{$(3,0)$} \\
\hline\hline
& $D_{n0}$~(m$^{-1}$) && 2.27 & 0.557 & 0.460 \\
\hline
&& SF$_6$ & 65.0 & 15.9 & 13.2 \\
&&     Xe & 58.3 & 14.3 & 11.8 \\
4.00 & $\displaystyle\frac{P(r_2)}{\epsilon_q\,Q\,V(\omega_{n0})}$ & Kr & 37.4 & 9.17 &7.57 \\
& (MPa/V) & Ar & 17.8 & 4.37 & 3.61 \\
&& N$_2$ & 12.5 & 3.06 & 2.52 \\
\hline\hline
& $D_{n0}$~(m$^{-1}$) && 1.87 & 0.311 & 0.298 \\
\hline
&& SF$_6$ & 99.4 & 16.5 & 15.8 \\
&&     Xe & 89.2 & 14.8 & 14.2 \\
6.35 & $\displaystyle\frac{P(r_2)}{\epsilon_q\,Q\,V(\omega_{n0})}$ & Kr & 57.2 & 9.51 & 9.11\\
& (MPa/V) & Ar & 27.2 & 4.53 & 4.34 \\
&& N$_2$ & 19.1 & 3.17 & 3.04 \\
\hline\hline\end{tabular}
\caption{\label{neures}Calculated results of acoustic effects due to charge asymmetry in spherical cavities having $r_2=20$~cm, for $r_1=4$~cm and $r_1=6.35$~cm. Dynamic pressure values are calculated through Eq.~(\ref{neupress}). All the numbers refer to $P_0=1$~atm and $T=298$~K -- see Eq.(\ref{delta}).}
\end{center}
\end{table}

In analogy with the case of electric polarization, the expression of the dynamic pressure peak values at $r_2$ due to the charge asymmetry is cast in the form
\begin{equation}
P(r_2,\nu_{n0})=D_{n0}\,Q_{n0}\,\epsilon_q\,M\,\delta\,V(\omega_{n0}).
\label{neupress}
\end{equation}
The results for the cavities considered in the present work are summarized in Tab.~\ref{neures}. If we suppose $\epsilon_q\sim10^{-21}$, $V\sim1$~kV and $Q\sim10^3$, a dynamic pressure $P\sim1\;\mu$Pa is expected.

A similar calculation performed on the first radial mode of the cavity of Ref.~\cite{Dylla1973} ($r_1=1.27$~cm and $r_2=19.7$~cm, 1~atm of SF$_6$) gives the following result:
\[\frac{P(r_2,\nu_{10})}{\epsilon_q\,Q_{10}\,V(\omega_{10})}=21.2\text{~MPa/V}\qquad\text{(exp.\ of Ref.~\cite{Dylla1973}),}\]
to be compared with the corresponding value that can be calculated using Eq.~(2) of the same reference, which is a factor about sixty larger ($P_D=1.24$~GPa/V). Also in this case we suspect that the neglecting by Dylla and King of the Bessel functions of the second kind in the solution of the wave equation might be the reason for the discrepancy.

\section{Experimental set-up and method}

\begin{figure}[htb]
\begin{center}
\includegraphics[width=8cm]{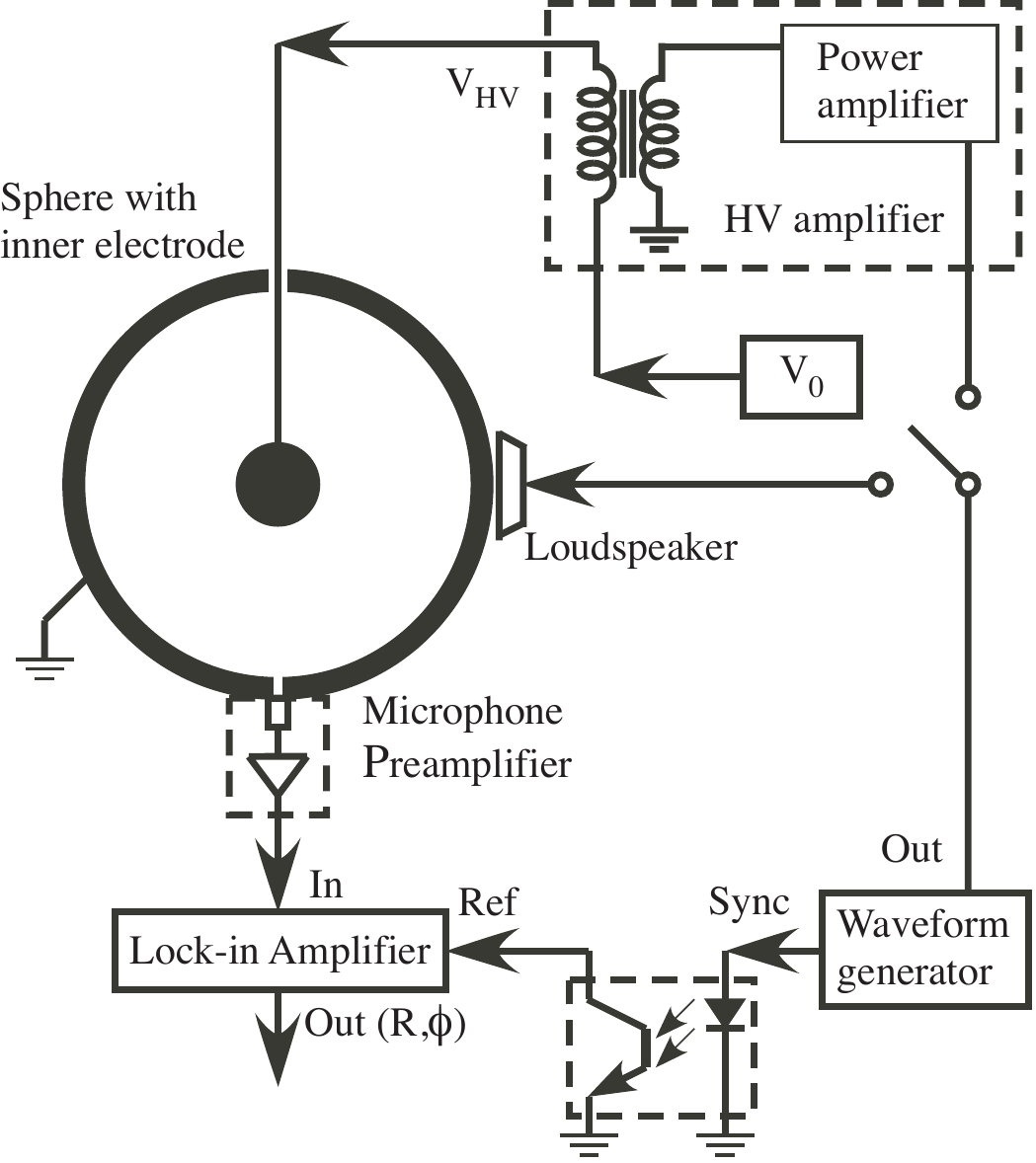}
\caption{\label{schema}Principle scheme of the experiment.}
\end{center}
\end{figure}
\begin{figure}[htb]
\begin{center}
\includegraphics[width=5cm]{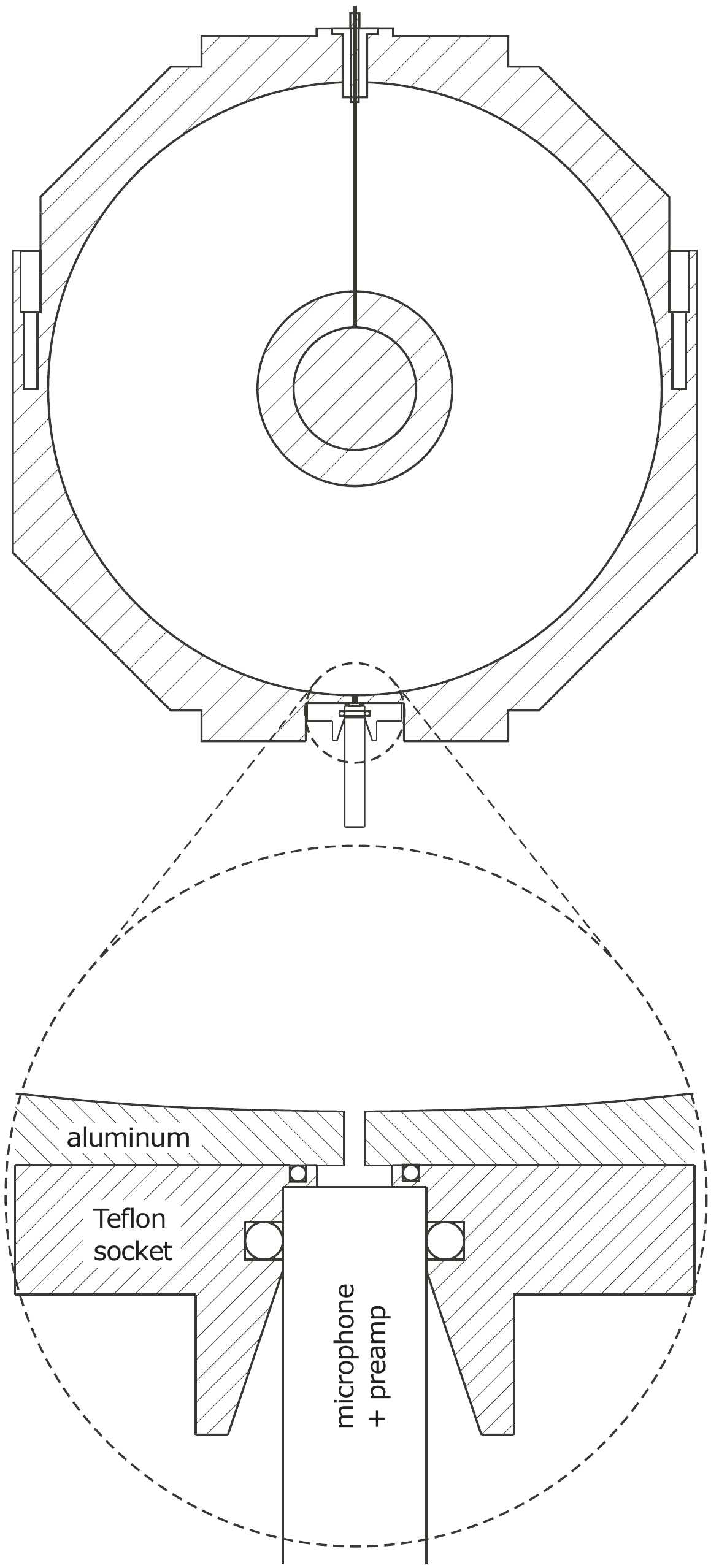}
\caption{\label{disegno}Top: schematic mechanical drawing of the cavity; the two different values of the internal electrode diameter are drawn. Bottom: detail of the microphone mounting.}
\end{center}
\end{figure}

A principle scheme of the experimental apparatus is shown in Fig.~\ref{schema}. The core of the set-up is a 40-cm diameter spherical cavity made of two mechanically joined hemispheres carved from a block of Al5056 aluminum alloy. The internal surface of the hemispheres is machined within a 0.1~mm precision and polished with $6-8\;\mu$m sandpaper. The thickness of the shell is not uniform, as roughly sketched in Fig.~\ref{disegno}.  Two holes are drilled in the sphere, a 1.6~cm diameter one located at the north pole and a 2~mm one at the south pole. In the course of the experiment the two holes played different roles: in a first phase, the upper hole was a gas inlet and was crossed by the suspension wire of the inner electrode; a bellows operated translation stage allowed the vertical positioning of the central electrode with a 1~mm resolution; the acoustic signal was recorded at the lower hole. Later, when there was evidence that the empty volume of the bellows was altering the amplitude of the resonant modes of the cavity due to the phenomenon of Helmholtz resonance \cite{Olson1967}, the height adjusting facility was abandoned and the upper hole was closed with an home made Macor electric feedthrough; the lower hole assumed then also the function of gas inlet. Two spherical internal electrodes have been used in turn: an 8~cm diameter aluminum one suspended to a 0.25~mm brass thread, and a 12.7~cm stainless steel one hanging by a 1~mm stainless steel wire. In order to provide acoustic and thermal isolation, the sphere is housed inside a large vacuum chamber where a rough vacuum ($\sim10^{-2}$~mbar) is made. Inside the chamber the sphere lays on a thick rubber pad. The apparatus is supported by a pneumatically damped optical table; by varying the relative pressure of the legs, the tilt angle could be controlled at the milliradiant level. The cavity can be evacuated down to $10^{-5}$~mbar and then filled with very high purity gases up to a pressure of 10~atm. The complete set-up lies in a temperature regulated room ($T\approx295$~K). After gas insertion, a time delay of at least one day was generally kept before taking measurements.

The outer shell is in electric contact with ground and isolated from the inner electrode. An HV feedthrough connects the inner electrode to an HV amplifier driven by a waveform generator; in this way an oscillating electric field is established inside the cavity. The HV amplifier is composed of a very high linearity power stage capable of up to 100~V$_\text{pp}$ output (5~V$_\text{pp}$ input) with an output impedance less than $1~\Omega$. The power amplifier is followed by a 1:100 transformer wound around an ultra linear M0 grade magnetic core, with total power 1.5~kVA. The whole HV amplifier is actually located in a different room than the sphere, inside a Faraday chamber. The ratio of the amplitudes of the second to the first harmonic has been measured to be less than $5\times10^{-5}$ at the amplifier output. Ac and dc components of the electric voltage $V_{HV}$ are measured using two six and a half digit multimeters. The ac readout is made through a 1:1000 voltage divider.

\begin{figure}[htb]
\begin{center}
\includegraphics[width=8cm]{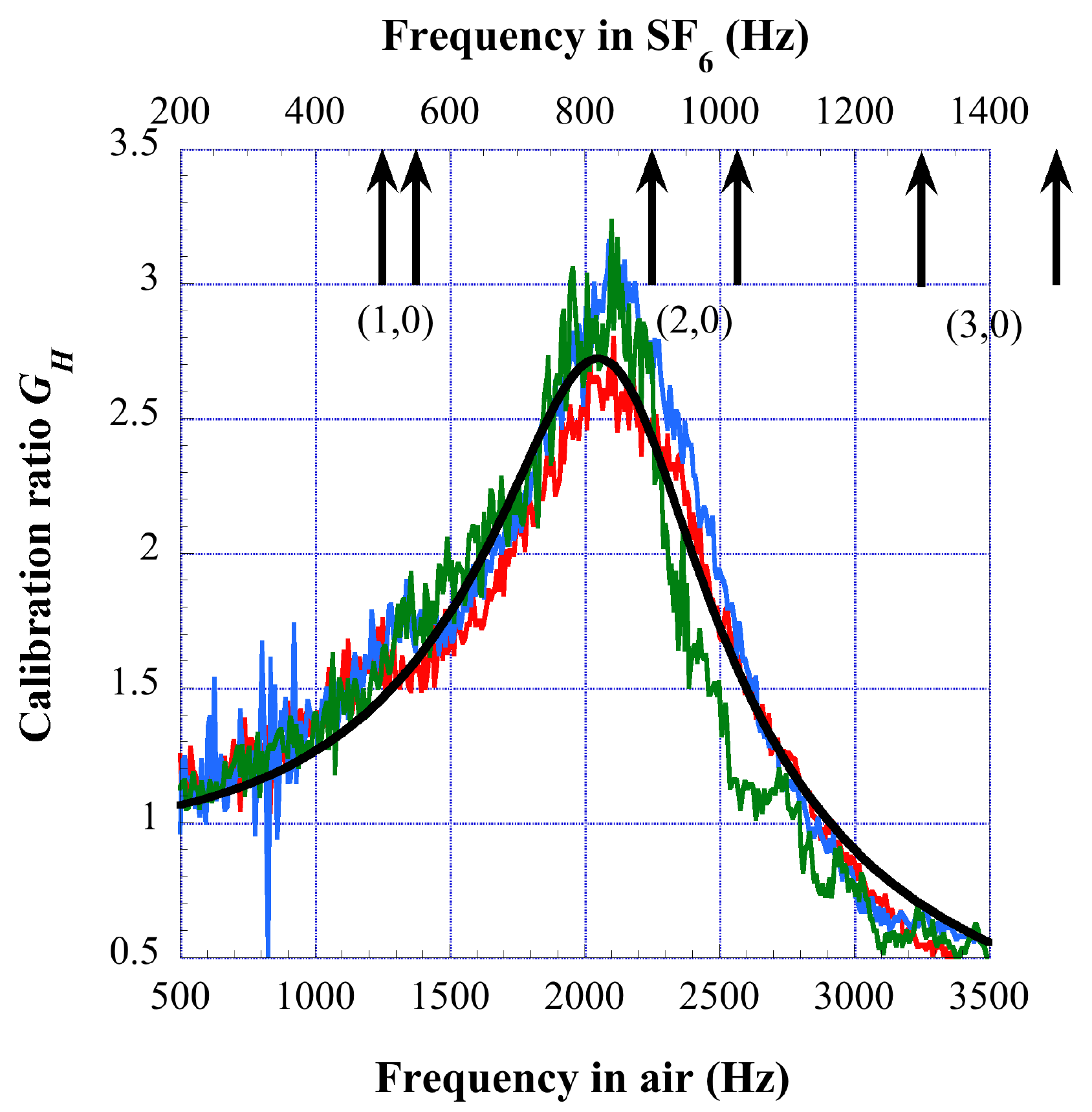}
\caption{\label{Helmcal}(Color online). Measured ratio $G_H$ of the frequency response in air of the microphone mounted as in Fig.~\ref{disegno}, to its free field response for three different distances from the loudspeaker: 25~cm, 35~cm and 58~cm. As the three graphs are quite similar, the smooth curve fits the average (not shown) of the three with the resonance function $y=A\left[\right(\nu^2-\nu_0^2\left)^2+\nu^2\Gamma^2\right]^{-1/2}$ with $A=4.58$~kHz$^2$, $\nu_0=2.13$~kHz, $\Gamma=0.80$~kHz. The arrows mark the positions of the first radial modes $(n,0)$ for the cavities with $r_1=4$~cm and $r_1=6.35$~cm.}
\end{center}
\end{figure}
\begin{figure}[htb]
\begin{center}
\includegraphics[width=7cm]{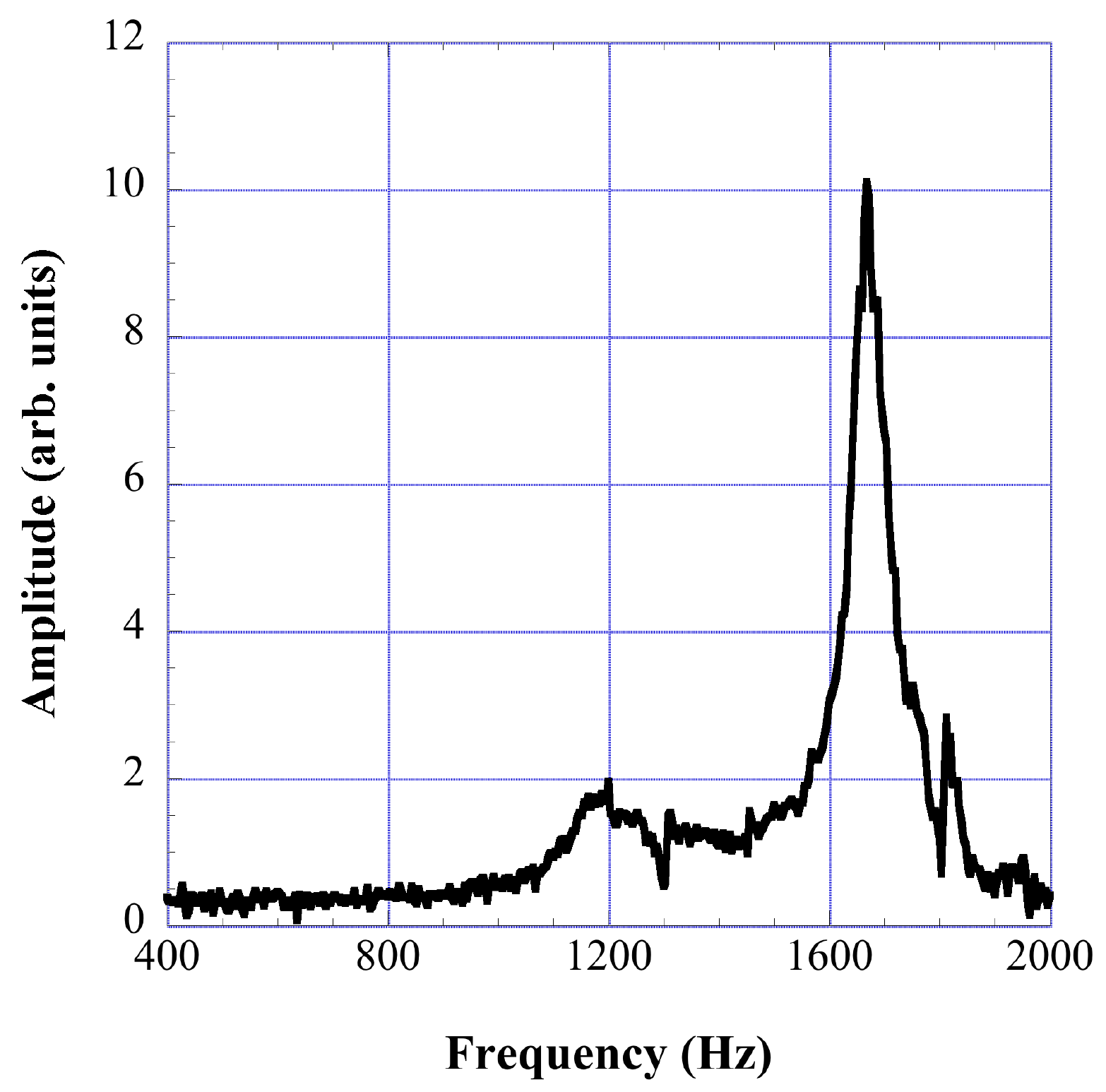}
\caption{\label{vibrspectr}Acoustic signal from the microphone due to mechanical vibrations of the evacuated cavity. A small loudspeaker is used as an excitation source.}
\end{center}
\end{figure}

As the microphone is sensitive to electric fields, it could not be flush mounted, namely with the sensitive membrane mimicking the inner surface of the cavity. In fact, we observed that in this last configuration an oscillating electric field generates a spurious acoustic signal which grows with the square of $E$. Acoustic oscillations of the gas inside the cavity are measured instead through a 2~mm diameter, 5~mm deep hole by means of an half inch prepolarized preamplified free-field condenser microphone (PCB 378B02); the head of the microphone is plugged inside a Teflon socket which lays against the outer surface of the sphere; an half-inch diameter Viton o-ring seals the contact between Teflon and aluminum (see lower part of Fig.~\ref{disegno}); the o-rings are not meant to stand any differential pressure, as the whole microphone assembly is in the same atmosphere of the cavity. The microphone is mounted on a linear translation stage; this measure became necessary since when the south pole hole was the only opening left in the sphere, to allow for vacuum pumping and gas filling the cavity.

Nominal microphone sensitivity is $K_m=50$~mV/Pa at 1~atm in the frequency interval 3.15~Hz~--~20~kHz; each microphone employed in the experiment had a calibration sheet, the actual value of the sensitivity being known with a 0.01~mV/Pa uncertainty. At higher (lower) pressures the microphone loses (gains) sensitivity at a rate of 0.013~dB/kPa, so that $K_m=27.3$~mV/Pa at 5~atm. The noise limit of the preamplifier is $1\;\mu\text{Pa}/\sqrt{\text{Hz}}$. The cavity and the duct in front of the microphone constitute an Helmoltz resonator, whose response has been measured in air and compared to the free field response of the microphone. Fig.~\ref{Helmcal} shows the results of this calibration: the Helmholtz cavity amplifies a factor $G_H\approx1.6$ at the first radial mode of the cavity with $r_1=6.35$~cm, much the same as the second radial mode, while the third radial mode is attenuated. The microphone is also sensitive to vibrations, hence it can be used to record a spectrum of the normal modes of vibration of the aluminum skin. In Fig.~\ref{vibrspectr} such a spectrum is shown; the data are obtained by mechanically exciting the evacuated aluminum cavity with an external source. As can be seen, no mechanical resonance appears below 1~kHz.

The microphone output is amplified and read using a dual phase Lock-In Amplifier (LIA) referenced to the waveform generator. A FFT spectrum analyzer is also used to analyze the microphone signal. To avoid electric pick-ups the LIA reference input is connected to the sync output of the signal generator through an optical link. This system allows pressure measurements of the type:
\begin{equation}
P(t) = P_1 \cos (\omega t + \phi),
\end{equation}
where the dynamic pressure amplitude $P_1$ and its phase $\phi$ are directly calculated from the (amplitude, phase) outputs of the LIA or of the FFT analyzer.

\begin{figure}[htb]
\begin{center}
\includegraphics[width=7cm]{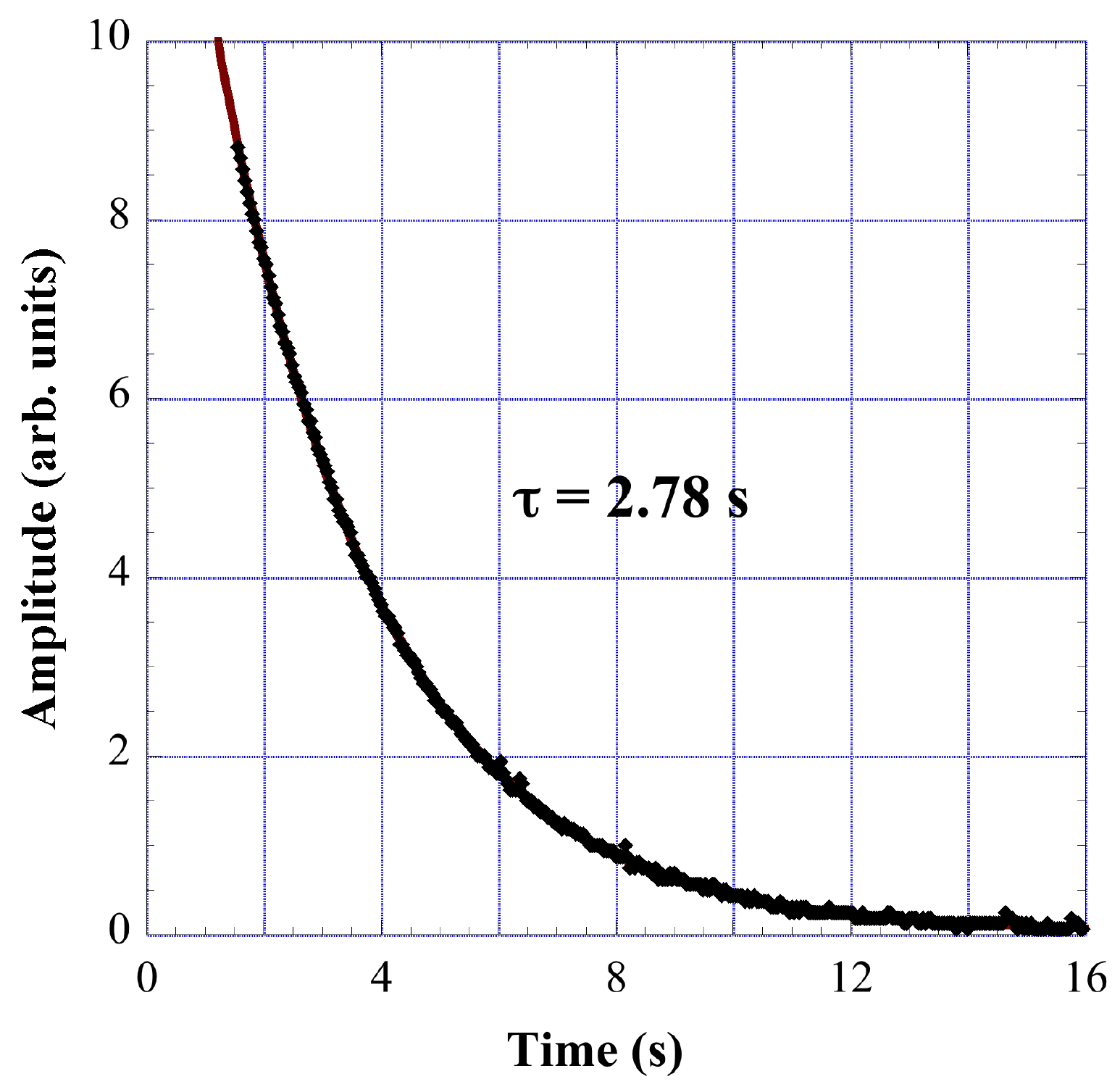}
\caption{\label{decayline}Decay of the demodulated microphone signal from the (1,0) resonance mode of the cavity with inner electrode radius $r_1=6.35$~cm filled with 1~atm of SF$_6$; the exponentially decaying fit line, completely masked by the experimental points, gives a quality factor $Q=\pi\nu_{10}\tau=4780$.}
\end{center}
\end{figure}
\begin{figure}[htb]
\begin{center}
\includegraphics[width=8cm]{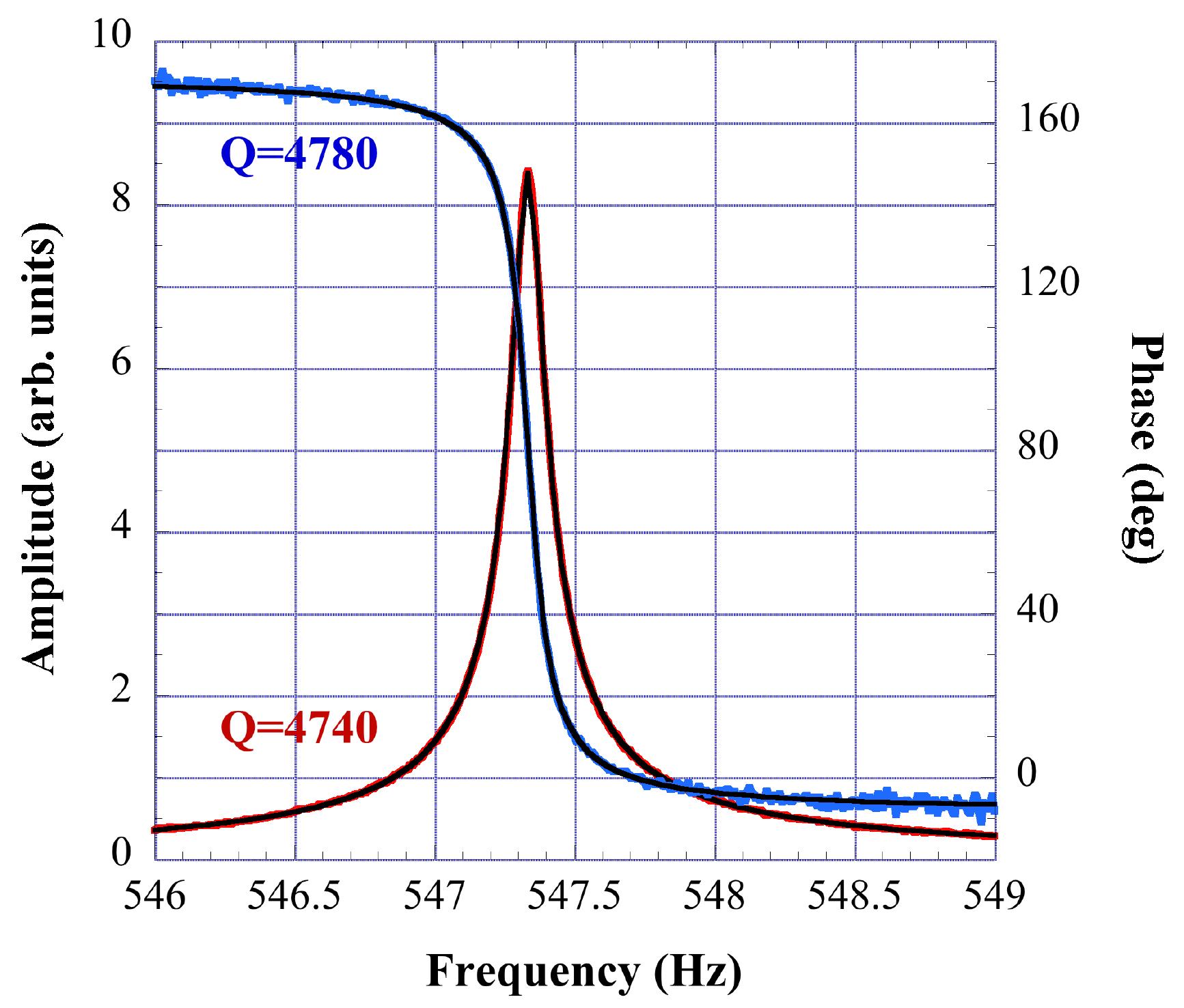}
\caption{\label{resline}(Color online). Lineshape and phase of the (1,0) resonance mode of the cavity with inner electrode radius $r_1=6.35$~cm filled with 1~atm of SF$_6$; the two lines are independently fit, giving compatible $Q$ values; the fit functions are completely masked by the experimental points.}
\end{center}
\end{figure}
\begin{figure}[htb]
\begin{center}
\includegraphics[width=8cm]{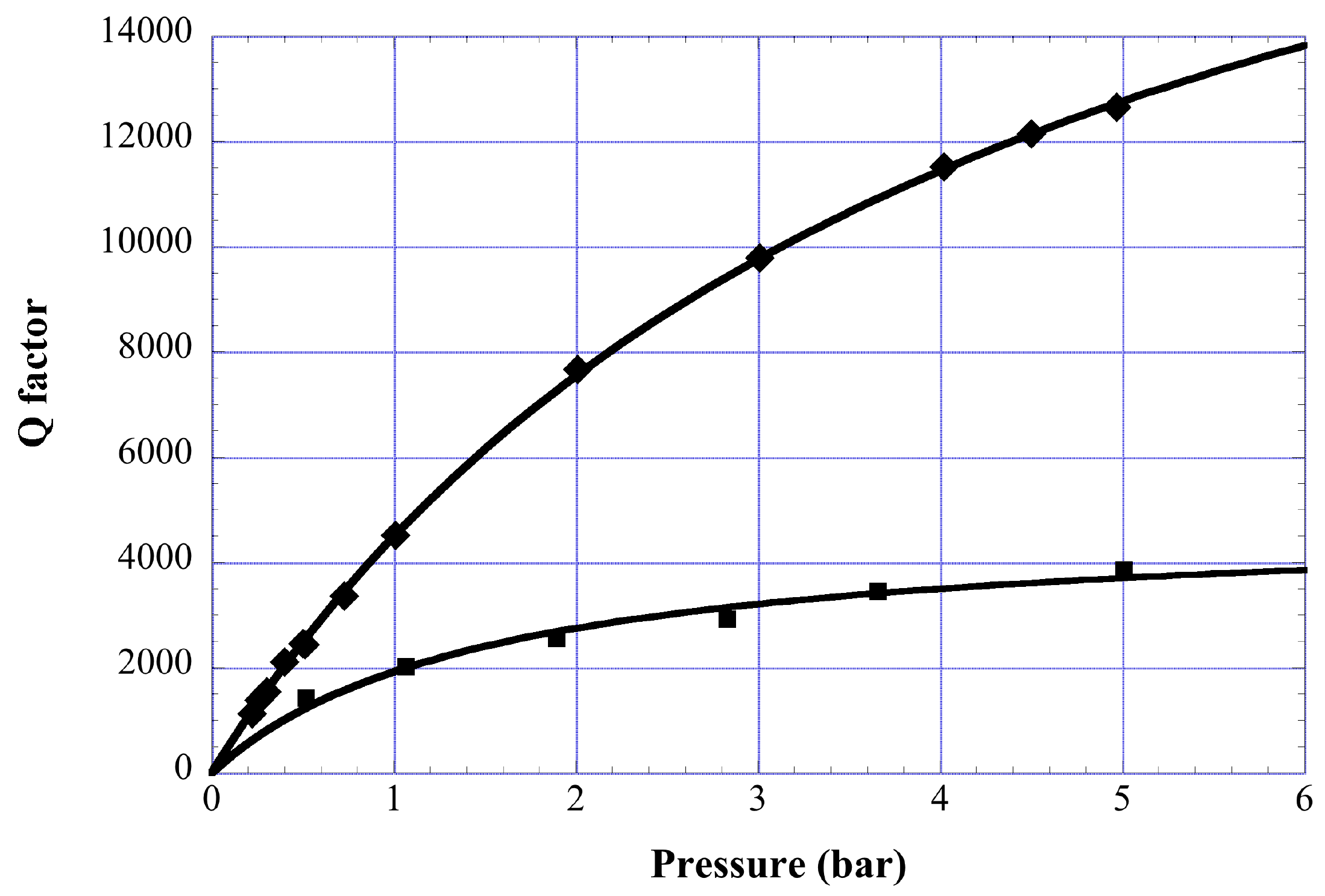}
\caption{\label{QvsP}Pressure dependence of the quality factor of the (1,0) resonance mode for SF$_6$ in the cavity with $r_1=6.35$~cm (upper curve, diamonds) and for Ar in the cavity with $r_1=4$~cm cavity (lower curve, squares). Continuous lines are only meant to guide the eye.}
\end{center}
\end{figure}
\begin{figure}[htb]
\begin{center}
\includegraphics[width=8cm]{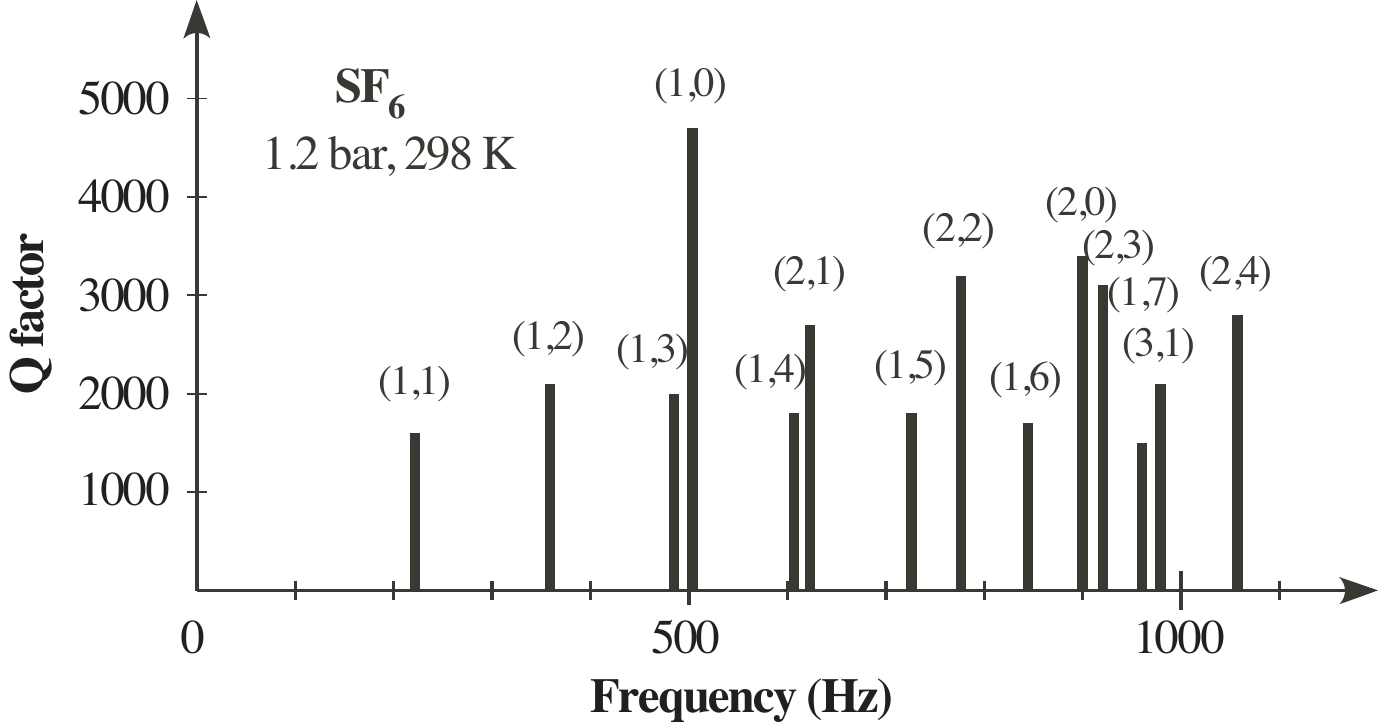}
\caption{\label{resspectr}Observed spectrum and $Q$ values of the cavity with $r_1=4$~cm, with 1.2~bar of SF$_6$.}
\end{center}
\end{figure}
\begin{figure}[htb]
\begin{center}
\includegraphics[width=8cm]{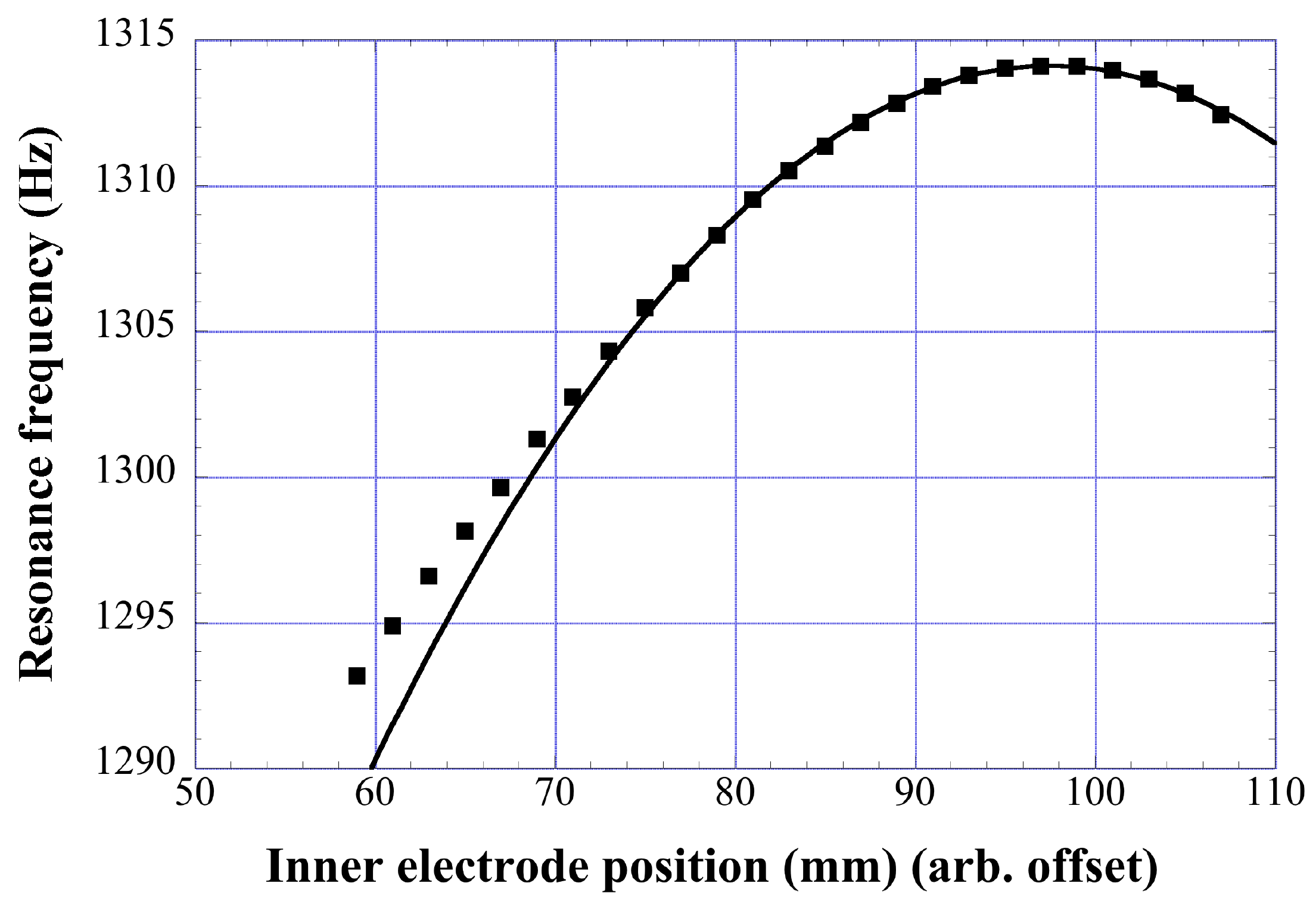}
\caption{\label{nuvsz}Resonance frequency of the $(1,0)$ mode of the $r_1=4$~cm cavity filled with 1~atm of N$_2$ as a function of the vertical position of the central electrode. Following Ref.~\cite{Kanellopoulos1978} the data are fitted with a parabolic curve.}
\end{center}
\end{figure}

In order to exploit the amplification factor of the acoustic modes, all the measurements have been performed in correspondence of the first cavity radial modes, having resonance frequencies $\nu_{n0}$. The acoustic modes of the cavity are probed exciting sound waves either using a loudspeaker external to the sphere or through electric polarization, by placing an oscillating high voltage on the inner electrode. The frequency of the resonant modes is determined maximizing the output of the LIA and monitored through its phase. The associated quality factor $Q$ is obtained from the decay time of the LIA output when the exciting source is switched off (see Fig.~\ref{decayline}). By scanning the exciting frequency, the lineshape of the resonance mode can also be traced directly from the output of the LIA or of the FFT (Fig.~\ref{resline}). The quality factor of a mode depends on pressure, increasing with increasing pressure (Fig.~\ref{QvsP}). The measured frequency spectrum of the cavity (Fig.~\ref{resspectr}) shows an excellent agreement with the calculated spectrum. Frequency shifts due to finite admittance of the cavity walls or to geometrical imperfections \cite{Mehl1982} are beyond the experimental sensitivity and the scope of the present work. The frequencies of the acoustic modes are seen to depend on the position of the central electrode. In Fig.~\ref{nuvsz} we have an example of this behavior: the resonance frequency shows a maximum which corresponds to a minimum in the eccentricity of the electrode \cite{Kanellopoulos1978}. 

\begin{table}[htb]
\begin{center}
\begin{tabular}{|l|c|c|c|c|c|c|}
\hline\hline
Measurement type & $\nu_{HV}$ & $V_0$ & $V_1$ & $P$ phase $\phi$ & $V^2(\omega_{n0})$ \\
\hline
Polarizability I (PI)   & $\nu_{n0}$   & $\neq0$ & $\neq0$ & $\phi_{HV}-\pi/2$  & $2V_0V_1$ \\
\hline
Polarizability II (PII) & $\nu_{n0}/2$ & 0       & $\neq0$ & $2\phi_{HV}-\pi/2$ & $V_1^2/2$ \\
\hline
Neutrality (N)          & $\nu_{n0}$   & 0       & $\neq0$ & $\phi_{HV}-\pi/2$  &   -----   \\
\hline\hline
\end{tabular}
\caption{\label{meastab}Experimental parameters for the different configurations of measure. The Fourier component of $V^2$ at a resonance frequency $\omega_{n0}$, $V^2(\omega_{n0})$, does not apply to neutrality measurements, which depend linearly on $V$.}
\end{center}
\end{table}

The measurements of interest are those connected with the presence of the electric field in the resonator. The high voltage signal on the inner electrode is parametrized as:
\[V_{HV}(t)=V_0+V_1\,e^{-i(\omega_{HV}t+\phi_{HV})}.\]
Three types of measurement are possible, depending on the values of $V_0$ and $V_1$; all the relevant values of the experimental parameters are summarized in Tab.~\ref{meastab}. Like in the original experiment by Dylla and King, the polarizability measurements are used to test and calibrate the apparatus. When the high voltage excitation frequency coincides with the frequency of a resonant mode and $V_0\neq0$, both effects due to polarization and to an hypothetical non neutrality of matter are present. In this case, considering the pressure signal as a function of $V_0$, one sees that the neutrality measurement N can be obtained as the limit for $V_0\rightarrow0$ of the polarization measurement PI. Moreover, this measurement configuration allows also a continuous monitoring of the forcing field, thus reducing phase drift effects.

\section{Results and discussion}

\subsection{Polarizability measurements}

\begin{figure}[htb]
\begin{center}
\includegraphics[width=8cm]{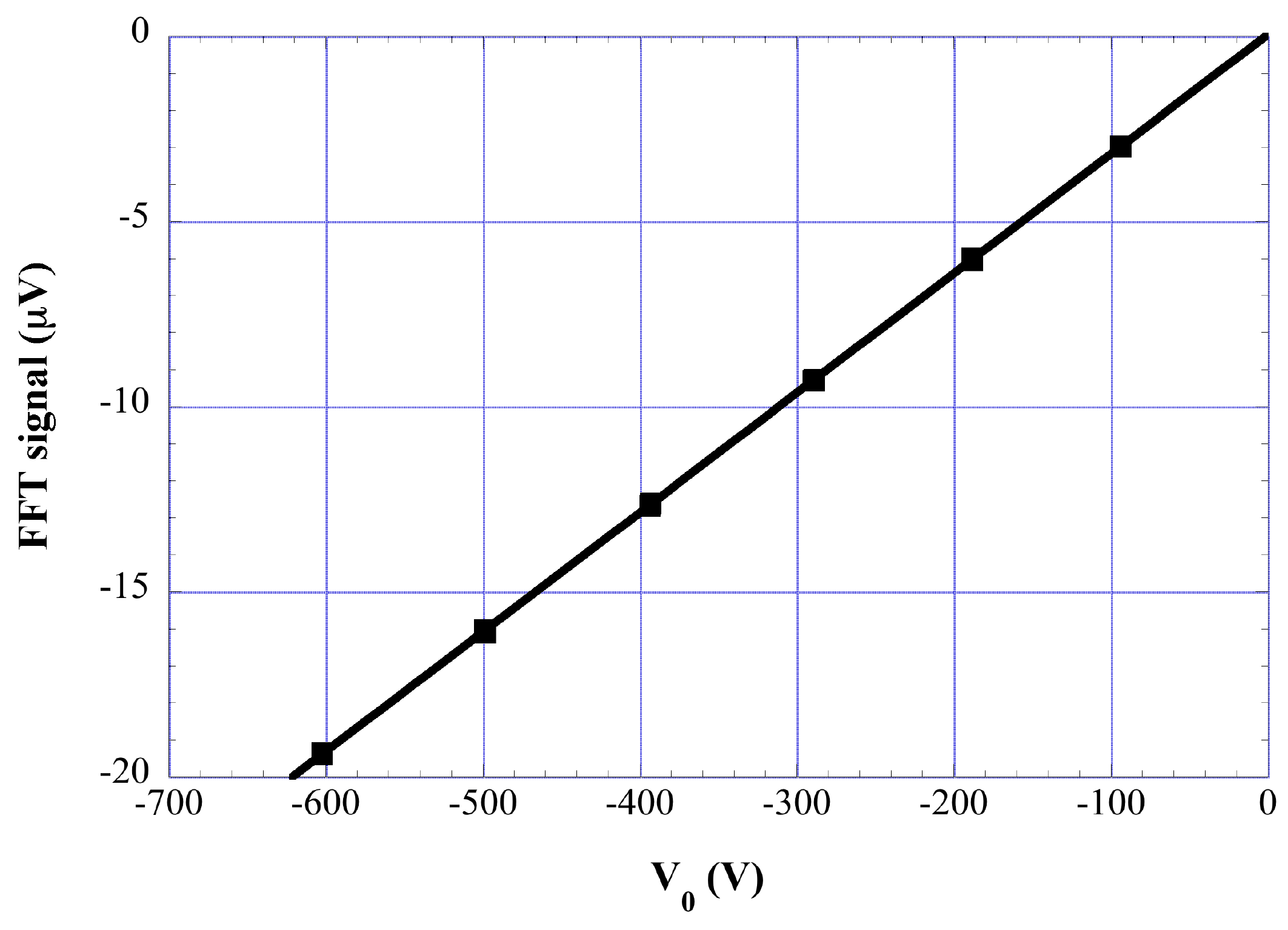}
\caption{\label{pol1}Typical PI-type polarization measurement for 0.25~bar of SF$_6$ on the (1,0) resonance mode at $\nu=551.6$~Hz of the cavity with $r_1=6.35$~cm, for $V_1=785$~V rms, as a function of $V_0$. All the 13 points drawn lay on a continuous straight line having slope $32.3\times10^{-9}$ and intercept 49.8~nV. The quality factor of the cavity is $Q=1400$. A 25~nV uncertainty on all the points results in a reduced $\chi^2$ equal to one.}
\end{center}
\end{figure}
\begin{figure}[htb]
\begin{center}
\includegraphics[width=8cm]{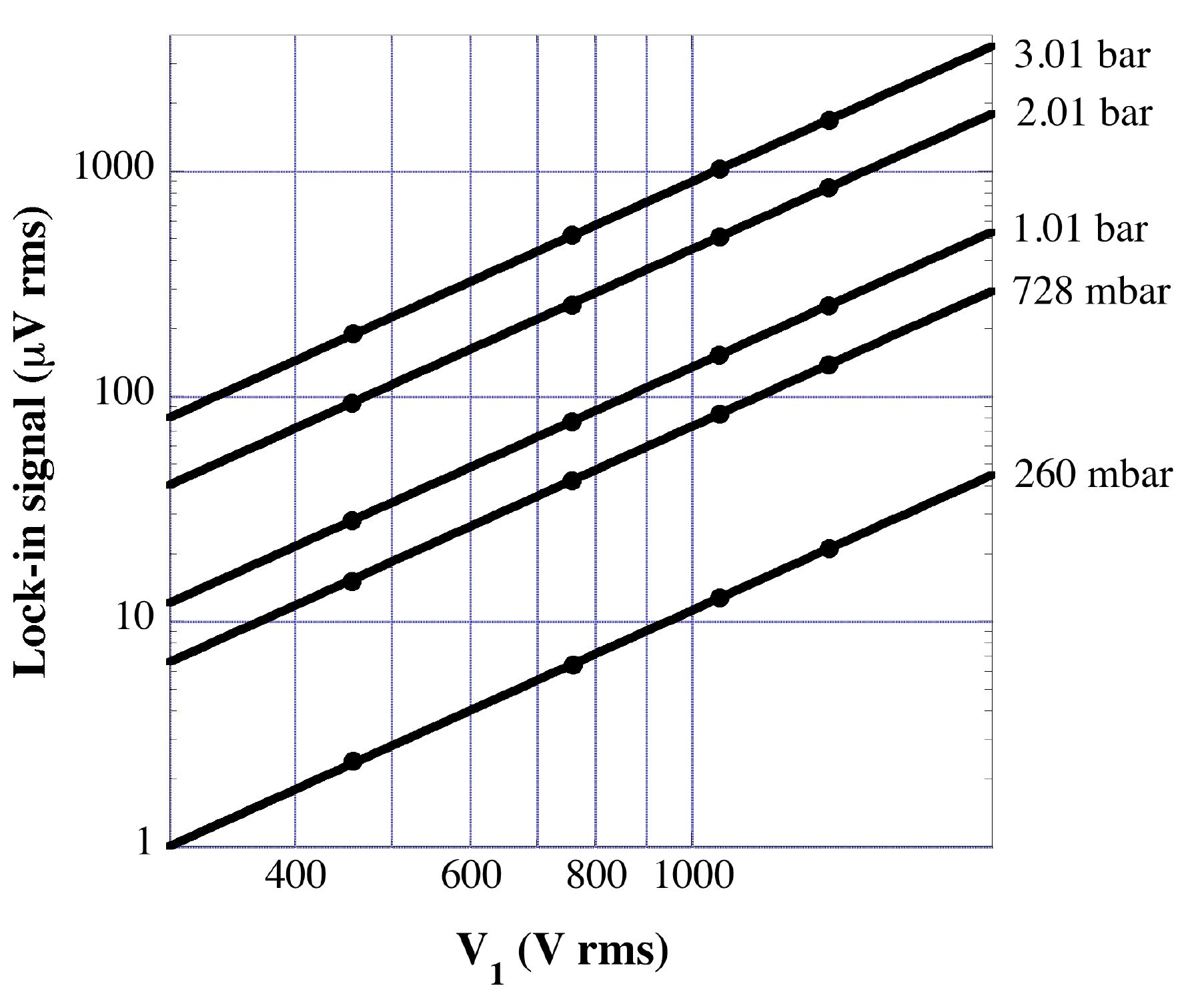}
\caption{\label{pol2}PII-type polarization measurements for several SF$_6$ pressures on the (1,0) resonance mode of the cavity with $r_1=6.35$~cm, as a function of $V_1$. The data are fitted with $V_1^2$ functions. Resonance frequencies and quality factors range from $\nu_{10}=550$~Hz and $Q=1400$ at $P_0=260$~mbar to $\nu_{10}=537$~Hz and $Q=9800$ at $P_0=3.01$~bar. Uncertainties between 20 and 50~nV result in reduced $\chi^2$'s equal to one.}
\end{center}
\end{figure}
\begin{figure}[htb]
\begin{center}
\includegraphics[width=7cm]{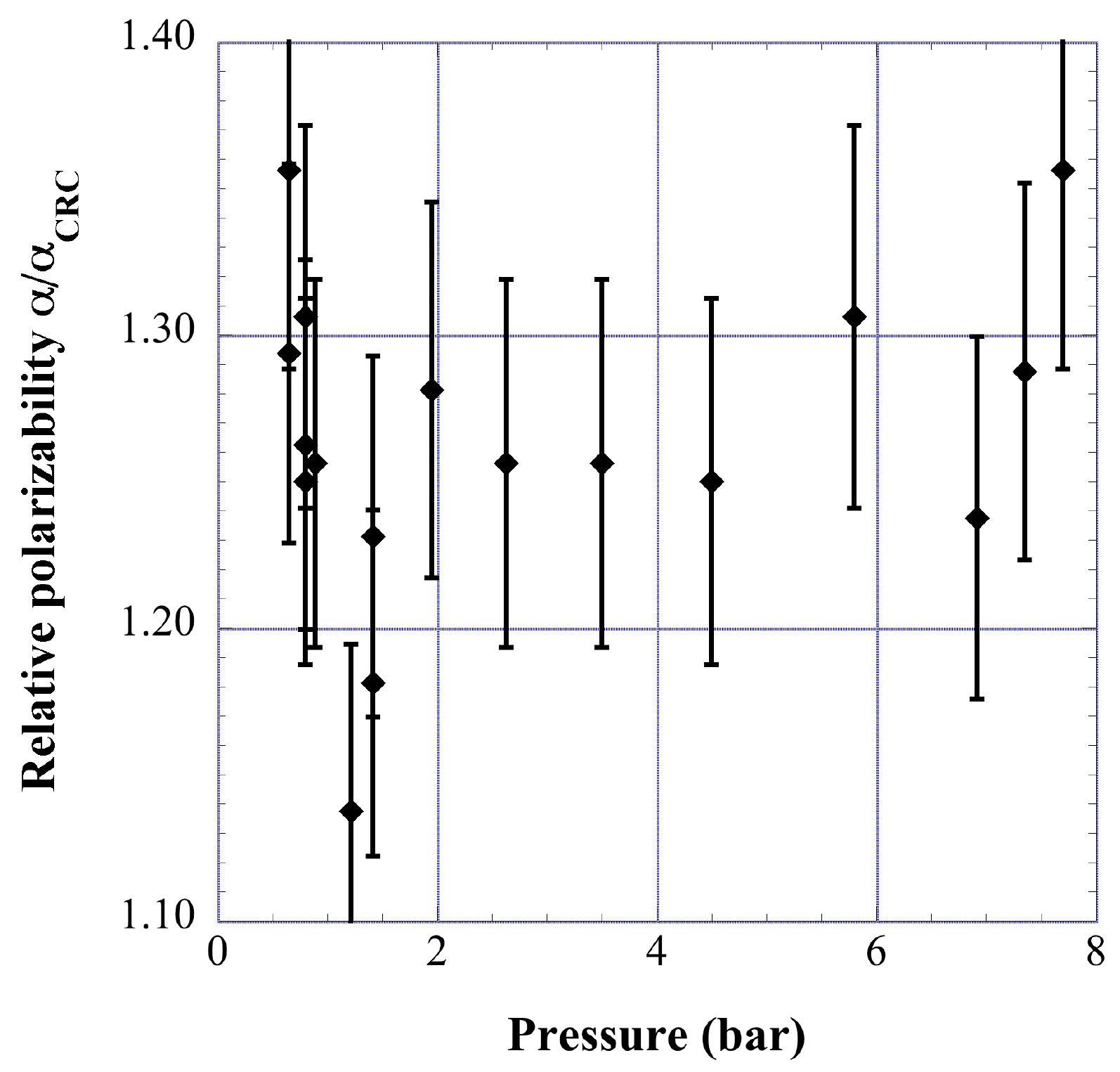}
\caption{\label{alphasf6}Measurements of the molecular polarization $\alpha$ of SF$_6$ as a function of pressure compared to the value $\alpha_\text{CRC}=7.277\;10^{-40}$~C$\cdot$m$^2$/V. In this graph the 0.013~dB/kPa correction of the sensitivity of the microphone has not been taken into account.}
\end{center}
\end{figure}

In Figs.~\ref{pol1} and \ref{pol2} typical results of the polarizability measurements are shown. In both cases no deviation of the data from the theoretical fit functions is observed. The slopes and curvatures obtained from the fits allow to calculate the electric polarizability of the gases. The data scale well with the gas pressure and with the resonance mode. A normalized plot of determinations of $\alpha$ in SF$_6$ at different pressures is shown in Fig.~\ref{alphasf6}; the measurements have been obtained from linear fits (PI-type), quadratic fits (PII-type), and from single measurements taken in a period of several months during which the apparatus underwent major changes that affected the uncertainty of the measurements but apparently did not alter the data distribution. As can be seen, the experimental value of $\alpha$ is a factor $R_\alpha\approx1.25$ larger than the tabulated one. The reasons for that are not clear. Note however that, since the polarization measurements represent a calibration of the apparatus at the working frequency, we have assumed that all the acoustic signals detected by the apparatus, including those measured in neutrality measurements, are sensed with a microphone sensitivity $K'_m=R_\alpha G_HK_m\approx100$~mV/Pa. In the following we consider two possible explanations for the discrepancy.

First of all, the gas is not isolated from the aluminum skin of the cavity. In fact, as mentioned before, a loudspeaker external to the cavity is able to excite the resonance modes of the gas. Conversely, during P-type measurements, an accelerometer in external contact with the cavity is able to detect the vibration of the wall of the cavity and decay curves of the type of Fig.~\ref{decayline} can even be recorded; no such effect can be observed if $\nu_{HV}$ is detuned from resonance, meaning that the coupling between the high voltage and the cavity skin is mediated by the resonating gas: the gas--skin system should then be considered a single resonating system. In the original experiment by Dylla and King, the cavity skin was a thin foil of copper, a less rigid material. A second possible source of the difference between calculation and measurement is the fact that the presence of the electric wire, to which the inner electrode is suspended, makes the electric field lines not perfectly radial; hence the acoustic and the electric fields do not completely match. Moreover, the highest values of the electric field are found near the wire.

We also performed measurements of the polarizability of other gases at 1~atm. For Xe and Kr the relative polarizability $\alpha_\text{rel}=\alpha/\alpha_\text{CRC}$ is similar to the value for SF$_6$: $\alpha_\text{rel}=1.29$ for Kr and $\alpha_\text{rel}=1.25$ for Xe. As the resonance frequency increases, we find $\alpha_\text{rel}=1.7$ for Ar and $\alpha_\text{rel}=1.55$ for N$_2$, due to the onset of the mechanical resonances (see Fig.~\ref{vibrspectr}).

\subsection{Neutrality measurements}

\begin{figure}[htb]
\begin{center}
\includegraphics[width=7cm]{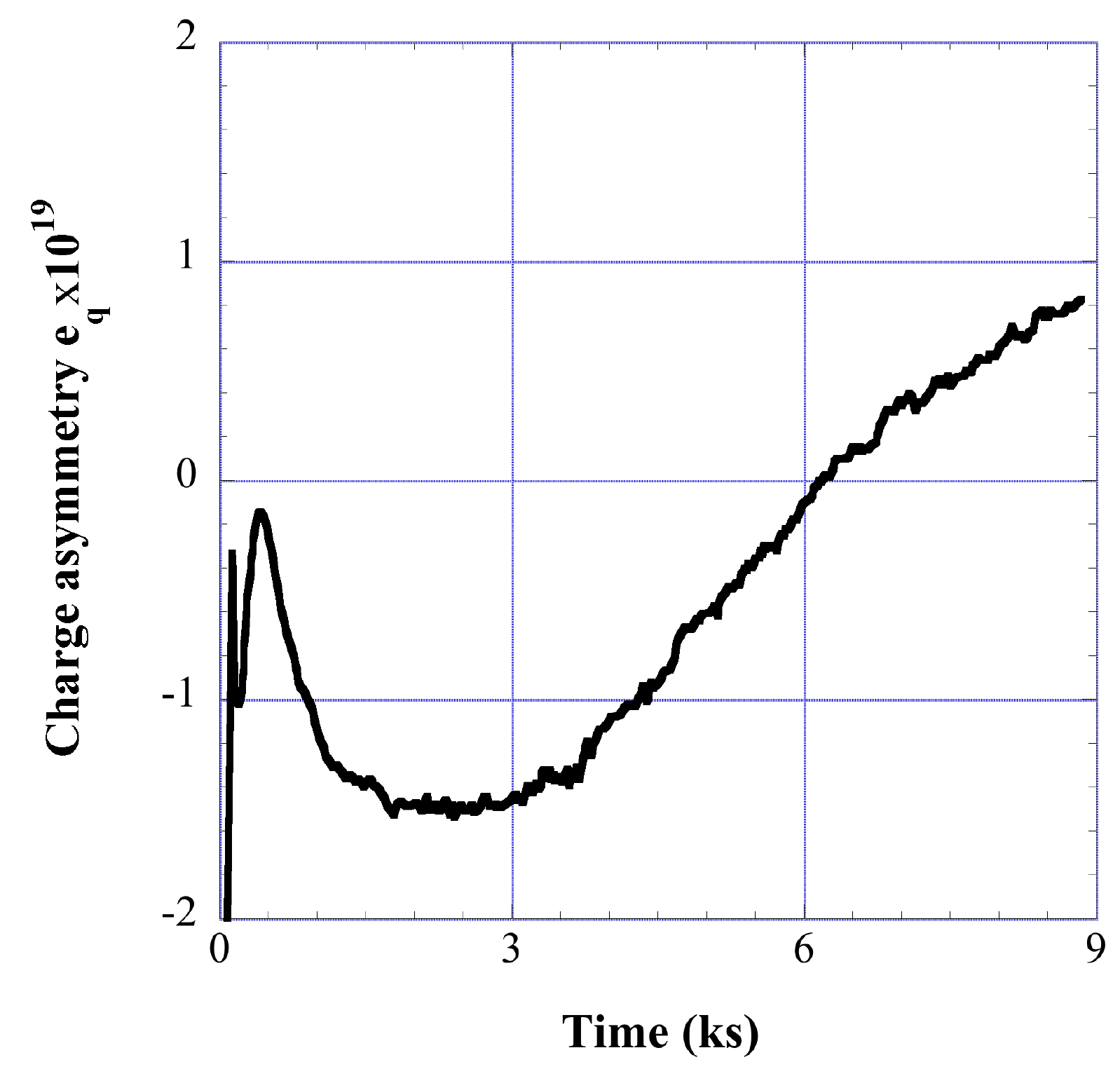}
\caption{\label{spurio}Demodulated dynamic pressure signal ($y$ or {\em in quadrature} output of the LIA), recorded in N-type measurement configuration with 0.48~bar of SF$_6$, following the application to the inner electrode of a dc high voltage $V_0=350$~V for ten minutes. After switching off the high voltage at $t=0$, $V_0$ decays with a time constant of about one minute. The $x$ or {\em in phase} output of the LIA is not shown, as during the observation time this last signal is fairly stable with values near zero.}
\end{center}
\end{figure}

As far as the neutrality measurements are concerned, no direct N-type measurement was possible. In fact, when $V_0=0$ and $\nu_{HV}=\nu_{n0}$, a small but significant spurious signal is always present, with the right phase and an amplitude corresponding to a non-neutrality of the matter of the order of $\epsilon_q\approx10^{-19}$. This signal is not an electric pick-up, as is proved for example by detuning the excitation frequency; moreover, it decays with the time constant of the cavity. It can have a mechanical contribution, due to waves excited by the high voltage in the skin of the cavity, an effect completely negligible as long as polarizability effects are concerned, but that can play a role in this context. A large contribution must however come from free charges in the gas. In fact, we observed an increase of the neutrality signal by irradiateing the cavity with $\gamma$-rays from a radioactive source. The spurious signal reacts with hysteresis to dc high voltages applied to the inner electrode for time periods of the order of a few minutes. A typical example of this behavior is shown in Fig.~\ref{spurio}. We have also observed that the spurious signal grows more than linearly with the base pressure inside the resonator. The pressure inside the resonator should then be kept as low as possible. With this finding, the idea of increasing the pressure inside the cavity to obtain larger signals reveals useless. None of these effects are reported in the 1973 paper by Dylla and King.

To go around the impossibility of a direct N-type measurement, we chose to use instead the intercepts in PI-type measurements taken at a base pressure $P_0\approx0.25$~bar of SF$_6$. As noted before, the intercepts represent a determination of the acoustic effect of an hypothetical charge asymmetry. We expected that the continuous presence of a DC voltage could sweep away the free charges from the gas. The hypothesis proved to be right: referring to Fig.~\ref{pol1}, where a typical set of measurements is shown, one can see that the value of the intercept is much smaller than that obtained in a direct N-type measurement. All the measurements have been taken on the (1,0) resonant mode at $\nu_{10}\simeq552$~Hz, with $Q=1400$. The data taking procedure is as follow: the slope and intercept of the straight line have been determined several times in a period of several weeks; each data set has about ten points with at least five different voltage values; the points have been sampled in a non monotone sequence with a few minutes between the two of them, in order to avoid transient phenomena; the positive and negative branches of each line have been sampled in different data sets, to keep constant the polarity of the dc voltage during each measurement. For each point the resonance frequency is tuned observing the phase of the polarization signal; the maximum error on the tuning of the resonance frequency is less than 4~mHz; for the 0.25~bar cavity this implies an uncertainty on the amplitude of 0.04\%. A linear regression was performed on each data set; all the points of each set were given the same statistical uncertainty chosen in such a way to have a reduced $\chi^2$ of the order of unity. The calculated statistical uncertainties lay within a factor two from the measured noise figure of the electronic chain.

\begin{figure}[htb]
\begin{center}
\includegraphics[width=8cm]{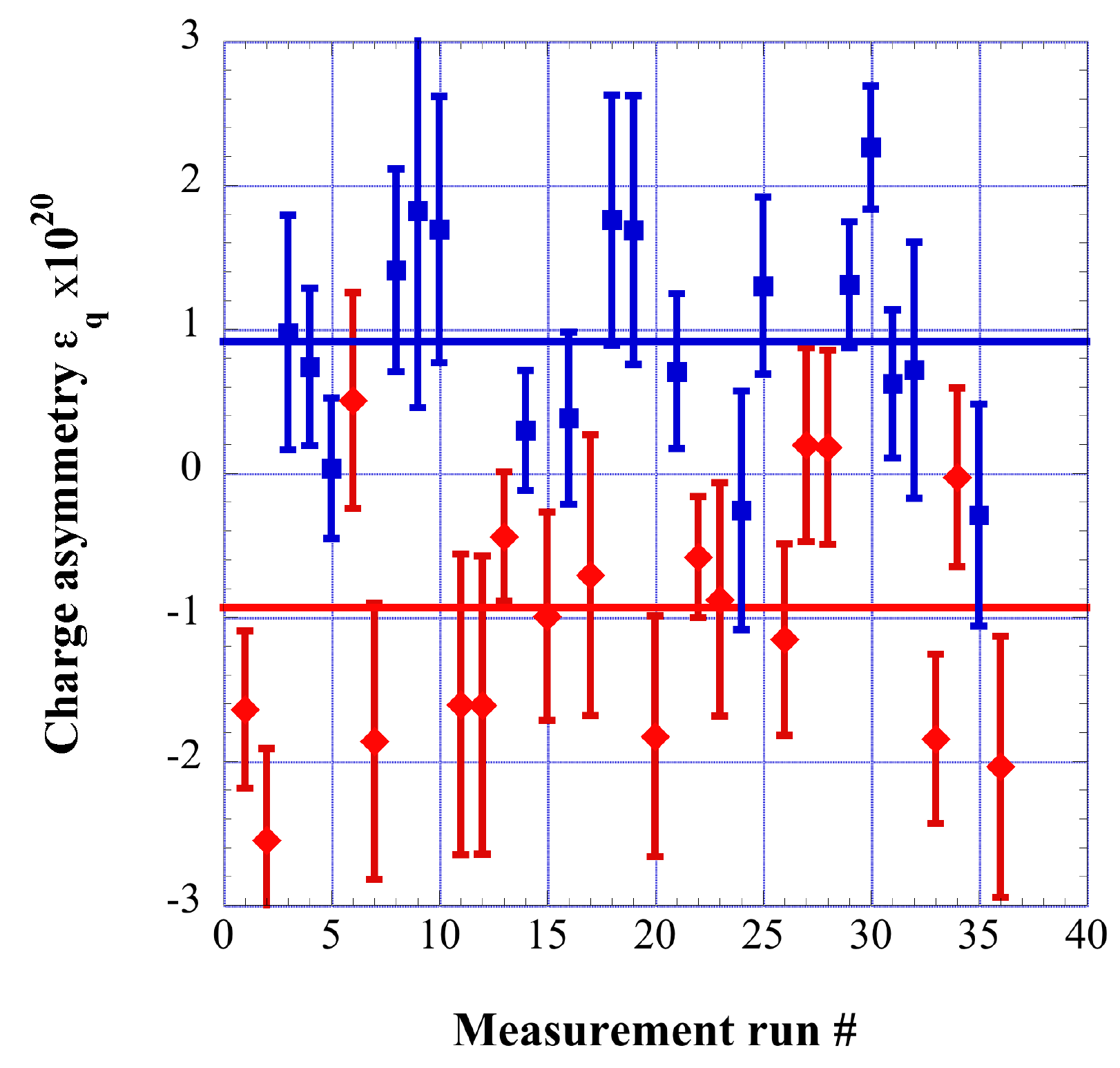}
\caption{\label{intercette}(Color online). Plot of thirtysix determinations of the charge asymmetry $\epsilon_q$ obtained from the intercepts of PI-type measurements taken on the resonance mode (1,0) of the cavity with $r_1=6.35$~cm with $P_0=0.25$~bar of SF$_6$. The factor $R_\alpha G_H\approx2$ (see Figs.~\ref{Helmcal} and \ref{alphasf6}) has been taken into account. The continuos lines are the averages, taken separately, of the data measured with negative (upper horizontal line, squares) and positive (lower horizontal line, diamonds) $V_0$. The two average values are $(+9.1\pm1.4)\,10^{-21}$ and $(-9.3\pm1.6)\,10^{-21}$, respectively.}
\end{center}
\end{figure}
\begin{figure}[htb]
\begin{center}
\includegraphics[width=6cm]{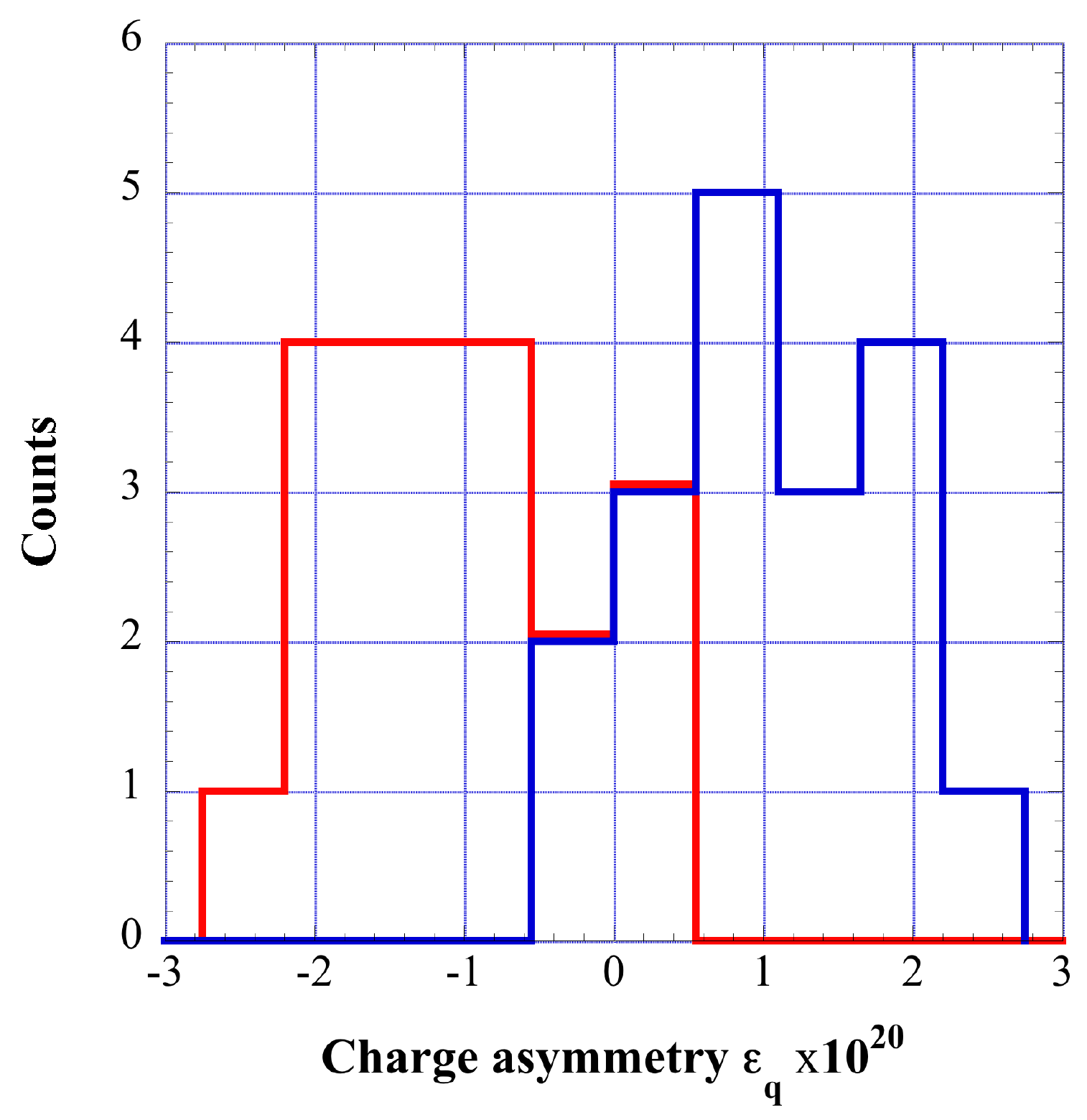}
\caption{\label{histogram}(Color online). Histograms of the two groups of data of Fig.~\ref{intercette}.}
\end{center}
\end{figure}

All the results obtained for the value of the intercept are shown in Fig.~\ref{intercette}. In Fig.~\ref{histogram} we show the histograms of the results. We note that not all the values are compatible with zero. It is observed that all but a few data sets taken with $V_0<0$ have positive intercept, while the opposite is true for the data sets taken with $V_0>0$. If the two groups of intercepts are separately averaged, opposite values are found: this means that besides the effect of polarization, at least another spurious effect of unknown origin must be present which, at variance with the hypothetical non neutrality of the matter, changes sign with the sign of the excitation signal. We get rid of this unwanted effect, thus extracting the value of the charge asymmetry, by making the half-sum of the two values. Finally, we find for the charge asymmetry of electron and proton as well as for the charge of the neutron
\[\epsilon_q=(-0.1\pm1.1)\,10^{-21}.\]
This result is compatible with zero and the sensitivity is at the level of the best results of Tab.~\ref{NeuTab}.

Improving this result depends little on having better statistics. Several other effects concur instead in frustrating further work on this set-up: the temperature stability of the whole apparatus, the temperature sensitivity of the voltage divider used in the readout of the ac component of $V_{HV}$, the sensitivity and linearity of the multimeter employed, the sensitivity of the microphone, the dynamic range of the spectrum analyzer. A better understanding of the spurious effects observed would also be necessary.

\section{Conclusions}

The acoustic technique of the 1973 experiment by Dylla and King \cite{Dylla1973} has been employed to test the neutrality of matter. A re-analysis  of Ref.~\cite{Dylla1973} shows that their claimed limit is invalidated by crucial mathematical errors. In our measurement no evidence of a charge asymmetry has been found. Assuming charge conservation in the $\beta$ decay, our result is at the level of the best existing limits, namely
\[\epsilon_q\lesssim1\times10^{-21}.\]
In the framework adopted, this limit holds both for the charge difference of proton and electron and for the charge of the neutron.

\begin{acknowledgments}
We thank E.~Berto for skilful mechanical help, G.~Galet for assistance with all the electric and electronic matters, M.~Romanato and M.~Zago for technical drawings and F.~Zatti for technical help. We thank Laboratori Nazionali di Legnaro for hosting the experiment and for support. We gratefully acknowledge advice and encouragement by Profs. F.~Borghesani and E.~Polacco and by Drs. A.~Bogi, I.~Pinto and N.~Stacchini of Acoustic Metrology SIT Center, ASL 7 Siena. We would also like to thank late Prof. N.~Cabibbo who stressed the importance of improving the existing limits on the neutrality of matter.
\end{acknowledgments}


\begin{thebibliography}{99}
\bibitem{CODATA2006}P.J.~Mohr, B.N.~Taylor and D.B.~Newell, Rev. Mod. Phys. \textbf{80}, 633 (2008).
\bibitem{Feinberg1959}G.~Feinberg and M.~Goldhaber, Proc. Natl. Acad. Sci. U.S.A. \textbf{45}, 1301 (1959).
\bibitem{Dirac1931}P.A.M~Dirac, Proc. R. Soc. A \textbf{{133}}, 60 (1931); Phys. Rev. \textbf{{74}}, 817 (1948); see also E.~Witten, Phys. Lett. \textbf{{86B}}, 283 (1979); S.Y.~Chu, Phys. Rev. Lett. \textbf{59}, 1390 (1987).
\bibitem{Okun1984}L.B.~Okun, M.B.~Voloshin and V.I.~Zakharov, Phys. Lett. \textbf{138B}, 115 (1984); R.~Foot {\em et al.}, Mod. Phys. Lett. A \textbf{5}, 1721 (1990); R.~Foot, H.~Lew and R.R.~Volkas, J. Phys. G \textbf{19}, 361 (1993).
\bibitem{Piccard1925}A.~Piccard and E.~Kessler, Arch. sci. phys. et nat. \textbf{7}, 340 (1925), cited in Ref.~\cite{Hughes1949}.
\bibitem{Hillas1959}A.M.~Hillas and T.E.~Cranshaw, Nature \textbf{184}, 892 (1959).
\bibitem{King1960}J.G.~King, Phys. Rev. Lett. \textbf{5}, 562 (1960); in a note to Tab.~1 of Ref.~\cite{Stover1967} inconsistencies in the results are reported.
\bibitem{Hughes1949}V.W.~Hughes, Phys. Rev. \textbf{76}, 474 (A) (1949); \textbf{105}, 170 (1957).
\bibitem{Lyttleton1959}R.A.~Lyttleton and H.~Bondi, Proc. R. Soc. A \textbf{252}, 313 (1959).
\bibitem{Zorn1960}J.C.~Zorn, G.E.~Chamberlain and V.W.~Hughes, Bull. Am. Phys. Soc. \textbf{5}, 36 (A) (1960).
\bibitem{Zorn1963}J.C.~Zorn, G.E.~Chamberlain and V.W.~Hughes, Bull. Am. Phys. Soc. \textbf{6}, 63 (A) (1961); Phys. Rev. \textbf{129}, 2566 (1963).
\bibitem{Fraser1968}L.J.~Fraser, E.R.~Carlson and V.W.~Hughes, Bull. Am. Phys. Soc. \textbf{13}, 636 (A) (1968).
\bibitem{Hughes1988}V.W.~Hughes, L.J.~Fraser and E.R.~Carlson, Z. Phys. D \textbf{10}, 145 (1988).
\bibitem{Shapiro1956}I.~Shapiro and V.~Estulin, Sov. Phys. JETP \textbf{3}, 626 (1956).
\bibitem{Shull1967}C.G.~Shull, K.W.~Billman and F.A.~Wedgwood, Phys. Rev. \textbf{153}, 1415 (1967).
\bibitem{Gahler1982}R.~G\"ahler, J.~Kalus and W.~Mampe, Phys. Rev. D \textbf{25}, 2887 (1982).
\bibitem{Baumann1988}J.~Baumann {\em et al.}, Phys. Rev. D \textbf{37}, 3107 (1988).
\bibitem{Trischka1960}J.W.~Trischka and T.I.~Moran, Bull. Am. Phys. Soc. \textbf{5}, 298 (A) (1960); C.~Becchi, G.~Gallinaro and G.~Morpurgo, Nuovo Cimento \textbf{39}, 409 (1965); V.B.~Braginskii, JETP Lett. \textbf{3}, 44 (1966); G.~Gallinaro and G.~Morpurgo, Phys. Lett. \textbf{23}, 609 (1966).
\bibitem{Stover1967}R.W.~Stover, T.I.~Moran and J.W.~Trischka, Phys. Rev. \textbf{164}, 1599 (1967).
\bibitem{Gallinaro1977}M.~Marinelli and G.~Morpurgo, Phys. Rep. \textbf{85}, 161 (1982); the quoted limit is a reassessment one order of magnitude worse of the results of G.~Gallinaro, M.~Marinelli and G.~Morpurgo, Phys. Rev. Lett. \textbf{38}, 1255 (1977).
\bibitem{Marinelli1984}M.~Marinelli and G.~Morpurgo, Phys. Lett. \textbf{137B}, 439 (1984).
\bibitem{Dylla1973}H.F.~Dylla and J.G.~King, Phys. Rev. A \textbf{7}, 1224 (1973).
\bibitem{PDG2010}See for example K.~Nakamura {\em et al.} (Particle Data Group), J. Phys. G \textbf{37}, 075021 (2010), p.~1135.
\bibitem{Unnikrishnan2004}C.C.~Unnikrishnan and G.T.~Gillies, Metrologia \textbf{41}, S125 (2004); this paper is also a fairly complete review on the topic of the neutrality of matter.
\bibitem{Arvanitaki2008}A.~Arvanitaki {\em et al.},
Phys. Rev. Lett. \textbf{100}, 120407 (2008).
\bibitem{Sengupta2000}S.~Sengupta, Phys. Lett. B \textbf{484}, 275 (2000); C.~Caprini and P.G.~Ferreira, J. Cosmol. Astropart. Phys. 02, 006 (2008).
\bibitem{Hughes1992}R.J.~Hughes and B.I.~Deutch, Phys. Rev. Lett. \textbf{69}, 578 (1992).
\bibitem{Bernstein1963}J.~Bernstein, M.~Ruderman and G.~Feinberg, Phys. Rev. \textbf{132}, 1227 (1963); G.~Barbiellini and G.~Cocconi, Nature \textbf{329}, 21 (1987); C.~Sivaram, Prog. Theor. Phys. \textbf{82}, 215 (1989); S.~Sengupta and P.B.~Pai, Phys. Lett. B \textbf{365}, 175 (1996); G.G.~Raffelt, Phys. Rep. \textbf{320}, 319 (1999).
\bibitem{Bondi1959}H.~Bondi and R.A.~Lyttleton, Nature \textbf{184}, 974 (1959); A.M.~Hillas and T.E.~Cranshaw, Nature \textbf{186}, 459 (1960).
\bibitem{King1973}J.G.~King, unpublished, reported in Ref.~\cite{Dylla1973}.
\bibitem{Stratton1941}J.A.~Stratton, {\em Electromagnetic Theory}, (Mac Graw-Hill, New York, 1941).
\bibitem{Trusler1991}J.P.M.~Trusler, {\em Physical Acoustics and Metrology of Fluids}, (Adam Hilger, Bristol, 1991).
\bibitem{Morse1953}P.M.~Morse and H.~Feshbach, {\em Methods of Theoretical Physics}, (McGraw-Hill, New York, 1953).
\bibitem{Ferris1952}H.G.~Ferris, J. Acoust. Soc. Am. \textbf{24}, 57 (1952).
\bibitem{CRC1995}D.R.~Lide ed., {\em CRC Handbook of Chemistry and Physics} 89$^\text{th}$ Ed., (CRC Press/Taylor and Francis, Boca Raton, FL, 2009).
\bibitem{Cazzola}P.~Cazzola and G.~Sartori, unpublished.
\bibitem{Olson1967}H.F.~Olson, {\em Music, Physics and Engineering}, (Dover, New York, 1967).
\bibitem{Mehl1982}J.B.~Mehl, J. Acoust. Soc. Am. \textbf{71}, 1109 (1982); {\em ibid.} \textbf{78}, 782 (1985); M.R.~Moldover, J.B.~Mehl and M.~Greenspan, {\em ibid.} \textbf{79}, 253 (1986); J.B~Mehl, {\em ibid.} \textbf{79}, 278 (1986).
\bibitem{Kanellopoulos1978}J.D.~Kanellopoulos and J.G.~Fikioris, J. Acoust. Soc. Am. \textbf{64}, 286 (1978).
\end{thebibliography}
\end{document}